\documentclass[prb,aps,twocolumn,showpacs]{revtex4}

\usepackage{graphics}
\usepackage{graphicx}
\usepackage{amssymb,amsmath}
\usepackage{bm}

\newcommand{\beq}{\begin{equation}}
\newcommand{\eeq}{\end{equation}}
\newcommand{\bea}{\begin{eqnarray}}
\newcommand{\eea}{\end{eqnarray}}

\begin{document}

\title{Quantum criticality in Kondo quantum dot coupled to helical edge 
states of interacting 2D topological insulators}
\author{ Chung-Hou Chung and Salman Silotri}
\affiliation{
Electrophysics Department, National Chiao-Tung University,
HsinChu, Taiwan, 300, R.O.C.
 }
\date{\today}

\begin{abstract}
We investigate  
theoretically the quantum phase transition (QPT) between the one-channel Kondo 
(1CK) and two-channel Kondo (2CK) fixed points in a quantum dot coupled 
to helical edge states 
of interacting 2D topological insulators (2DTI) 
with Luttinger parameter $0<K<1$.   
The model has been studied in Ref.\cite{TKNg}, and was 
mapped onto an anisotropic two-channel Kondo model via bosonization. 
For $K<1$, the strong coupling 2CK fixed point was argued to be stable 
for infinitesimally weak tunnelings between dot and the 2DTI 
based on a simple scaling dimensional analysis\cite{TKNg}.  
We re-examine this model beyond the bare scaling dimension analysis 
via a 1-loop renormalization group (RG) approach combined with bosonization 
and re-fermionization techniques near weak-coupling and strong-coupling 
(2CK) fixed points. We find for $K\rightarrow 1^{-}$ 
that the 2CK fixed point can be unstable 
towards the 1CK fixed point and the system may undergo a 
quantum phase transition between 1CK and 2CK 
fixed points. The QPT in our model comes as a result of 
the combined Kondo and the helical Luttinger physics in 2DTI, and it 
serves as the first example of the 1CK-2CK QPT that is accessible by 
the controlled RG approach. We extract quantum critical 
and crossover behaviors from various thermodynamical 
quantities near the transition. Our results are robust against 
particle-hole asymmetry for $\frac{1}{2}<K<1$. 
\end{abstract}

\maketitle

\section{ Introduction.}
Quantum phase transitions (QPTs)\cite{QPT}, the continuous phase 
transitions at zero temperature due to competing quantum ground states 
or quantum fluctuations, in correlated electron systems 
are of great fundamental importance and 
have been intensively studied over the past decades.
Very recently, nano-systems (in particular quantum dots\cite{QD})   
offer an excellent playground to study QPTs due to 
high tunability\cite{hur, furusaki, chungDQD, hofstetter, affleck,  
chung-noneq, kirchner, mitra}. 
The well-known Kondo effect\cite{hewson,Kondo} plays a crucial role 
in understanding low energy properties in quantum dot devices. 
Potential new QPTs in these systems may be realized in 
connection to exotic Kondo ground states. 
An outstanding example of 
an exotic Kondo state is  
the two-channel Kondo (2CK) system\cite{2CK, Affleck}, 
which has attracted much attention both 
theoretically and experimentally as it shows non-Fermi liquid behaviors 
at low temperatures. 
Experimentally, the 2CK behaviors have been realized 
 in Ref.\cite{goldhaber2CK} 
where a quantum dot independently couples to an infinite 
and a finite reservoirs of non-interacting conduction electrons.  

More interestingly, the 2CK physics has also been found theoretically 
in Kondo quantum dot 
coupled to two strongly 
interacting Luttinger liquid leads with Luttinger parameter 
$K<\frac{1}{2}$\cite{gogolin,kim}. In this case, electron-electron 
interactions in the leads strongly suppress the cross-channel Kondo 
correlations responsible for charge transport through the quantum dot 
while the Kondo correlations involving electrons on the same lead are 
unaffected, 
leading to an insulating two-channel Kondo ground state where 
two independent Kondo screenings occur between the spins on the dot 
and in each lead separately. On the other hand, for weaker electron 
interactions, $K>\frac{1}{2}$, both kinds of the 
Kondo scattering involving conduction 
electrons on the same and different leads become relevant at low energies, 
giving rise to conducting 1CK ground state where only a single-channel 
electrons (even combination of the two leads) effectively couple 
to the Kondo dot. An exotic quantum phase transition 
between the conducting 1CK phase for $K>\frac{1}{2}$ 
and the insulating 2CK behaviors for $K<\frac{1}{2}$ is therefore 
expected at $K=\frac{1}{2}$\cite{gogolin}. 
However, the critical properties of this 
1CK-2CK QPT have not yet been addressed yet since 
it is not accessible to any controlled theoretical approaches.    

On the other hand, recently 
a new type of materials--topological insulators (TIs)--with 
a gaped bulk and gapless edge states has been proposed 
theoretically\cite{TIs} and realized experimentally\cite{Hasan}. 
In 2D TIs, the gapless edge states have 
``helical'' nature, {\it i.e.} the directions of spin and momentum 
are locked together\cite{SCZhang}. Based on bosonization and a 
simple scaling dimension analysis, the 2CK behaviors were argued to 
be stabilized in Kondo quantum dot coupled to two interacting helical 
edge states of 2D TIs as long as a 
weak electron-electron interaction exists in the helical electrons  
($K<1$)\cite{TKNg}. However, 
on a general ground, similar competition between the cross-channel Kondo 
correlation and suppression of tunneling due to electron-electron 
interactions mentioned above is also expected here for the helical 
Luttinger liquid, a special type of Luttinger liquid with broken 
SU(2) symmetry. The exotic 1CK-2CK 
QPT may therefore occur in this new setup.

In this paper, we re-examine the system in Ref.\cite{TKNg} 
near 2CK fixed point to explore the possibility of the exotic 
1CK-2CK QPT via the controlled 
1-loop renormalization group (RG) approach combined with bosonization, 
which goes beyond the bare scaling dimension analysis in Ref.~\cite{TKNg}. 
For a weak but finite lead-dot tunneling, 
we find that for $K\rightarrow 1^{-}$ 
the 2CK fixed point can be unstable 
towards the (anisotropic) 1CK fixed point and the system may undergo a 
quantum phase transition (QPT) 
between 1CK and 2CK 
fixed points. The QPT in our model comes as a result of 
the combined Kondo and the helical Luttinger physics in 2DTIs. It 
serves as the first example of the 1CK-2CK QPT that is accessible by 
the controlled perturbative RG approach. The stability analysis 
on these two fixed points shows that they are stable against 
small particle-hole asymmetry for $\frac{1}{2}<K<1$.   
We extract the non-Fermi liquid behaviors from various thermodynamical 
quantities at the 1CK-2CK quantum critical point. 

This paper is organized as follows. In Sec. II, we introduce the model 
Hamiltonian and its bosonized form as shown in Ref.~\cite{TKNg}. In Sec. 
III. A.,  
we present the RG analysis of the model both in the weak coupling limit  
via bosonization approach (see Appendix A.). In Sec. III. B., we further 
map our bosonized 
model onto an effective Kondo model 
via re-fermionization near the strong-coupling 2CK fixed point. 
We then perform the RG analysis via both poor-man's scaling (see Appendix B.) 
and field-theoretical $\epsilon$-expansion technique 
(see Appendix C.). Our 
RG analysis in both limits suggests a quantum phase transition between 
1CK and 2CK fixed points.  In Sec. IV. we perform stability 
analysis on the 1CK and 2CK fixed points respectively. We 
find that both fixed points are stable for $\frac{1}{2}<K<1$, which 
substantiates our main finding that there exists an unstable quantum critical 
points separating two stable 1CK and 2CK fixed points near $K=1$. 
In Sec. V., we calculate via field-theoretical 
$\epsilon$-expansion approach the critical properties and crossover 
functions of various thermodynamical observables. In Sec. VI., we 
emphasize the clear physical picture of our main findings 
and draw conclusions.
\section{ Model Hamiltonian.}
In our set-up, the Kondo Hamiltonian has the same form as in 
Ref.~\cite{TKNg}, given by:
\begin{eqnarray}
H&=&H_{0}+ H_{K}+H_{int},\nonumber \\
H_{0}&=& -{\it i} v_F \sum_{i=1,2} \int_{-\infty}^{\infty} dx[c^{\dagger\uparrow}_{i,R}(x) \partial_x c_{i,R}^{\uparrow}(x)\nonumber \\
&-& c^{\dagger\downarrow}_{i,L}(x) \partial_x c_{i,L}^{\downarrow}(x)],\nonumber \\
H_K &=& \sum_{i=1,2}J_1 \vec{S}\cdot \vec{s}_{i,i}+\sum_{i\neq j} J_2 \vec{S}\cdot \vec{s}_{i,j},\nonumber \\
H_{int}&=&\sum_{i,\sigma=\uparrow\downarrow} g_4 \int [c^{\dagger\sigma}_{i,\alpha}(x) c_{i,\alpha}^{\sigma}(x)]^2 dx\nonumber \\
&+& g_2 \int c^{\dagger\uparrow}_{i,R}(x) c_{i,R}^{\uparrow}(x)c^{\dagger\downarrow}_{i,L}(x) c_{i,L}^{\downarrow}(x) dx.
\label{H}
\end{eqnarray}
Here, $H_0$ describes the two 
conduction electron baths (labeled as lead $1$ and lead $2$) 
made of helical edge states in 2D topological insulators, 
$H_K$ is the Kondo interaction, 
and the electron-electron interactions with forward scattering 
$g_{2,4}>0$ terms are given by $H_{int}$  
with $i=1,2$ the lead index, and $\alpha=R,L$ being the label of the 
right ($R$) and left ($L$) moving electrons in the helical edge state. 
The conduction electron spin operator is given by: 
$\vec{s}_{i,j} = \sum_{k,k^{\prime},\gamma,\delta}c^{\dagger\gamma}_{ki}\cdot 
\frac{\vec{\sigma}_{\gamma\delta}}{2}\cdot c_{k^{\prime}j}^{\delta}$ with 
$\gamma,\delta=\uparrow,\downarrow$, $i,j=1,2$. The local impurity spin operator 
on the quantum dot can be expressed in terms of pseudo-fermion operator 
$f_{\sigma}$\cite{coleman-costi}: 
$\vec{S} = \sum_{\gamma,\delta}f^{\dagger}_{\gamma}\cdot 
\frac{\vec{\sigma}_{\gamma\delta}}{2}\cdot f_{\delta}$. In the Kondo limit 
of our interest, the impurity (quantum dot) is singly-occupied: 
$\sum_{\sigma=\uparrow,\downarrow}f_\sigma^{\dagger}f_\sigma = 1$. 
Here, the couping $J_{1}$ and $J_2$ in $H_K$ stand for the strength of the 
Kondo correlations between the dot and electrons on the same and 
different leads, respectively. Note that in the presence of 
spin-orbit coupling, the spins $\uparrow$/$\downarrow$  
of the helical edge state electrons are locked 
with their right-moving ($R$)/left-moving ($L$) momentum. 
Note also that the small 
spin-orbit coupling will break the SU(2) spin-rotational symmetry 
 in the above isotropic Kondo model, leading to  the 
anisotropic Kondo model with $J_i^{xy}\neq J_i^z$\cite{TKNg}. 

The Hamiltonian Eq.~\ref{H} can be bosonized through 
the standard Abelian bosonization\cite{bosonization} for  
the electron operator\cite{TKNg,Kane}: $c_{i,R/L} = \frac{1}{\sqrt{2\pi a}} 
F_{i,R/L}e^{\pm {\it i}(\sqrt{4\pi} \phi_{iR/L}(x) + k_F x)}$; 
the bosonic fields $\phi_{i}(x) = \phi_{iL}(x) + \phi_{iR}(x)$, 
$\theta_{i}(x) = \phi_{iL}(x) - \phi_{iR}(x)$. The dual fields 
$\phi_i(x)$ and $\theta_i(x)$ obey the commutation relations:
$[\phi_i(x), \theta_j(x^{\prime})] = \frac{-{\it i}}{2}\delta_{ij} 
sgn(x-x^{\prime})$ with $sgn(x=0)=0$. The 
symmetric and antisymmetric combinations of $\phi_i$, $\theta_i$ 
are defined as: $\phi_{s/a}= \frac{1}{\sqrt{2}} (\phi_1 \pm\phi_2)$ and 
$\theta_{s/a}= \frac{1}{\sqrt{2}} (\theta_1 \pm \theta_2)$.  
Here, $F_{i, R/L}$ are the Klein factors to preserve the anti-commutation 
relations between fermions (electrons) in the bosonsized form, and $a$ is the 
lattice constant (lower bound in length scale); 
we have also dropped the spin indices of the edge state electrons 
due to their helical nature. 
The bosonized Hamiltonian after rescaling the boson fields 
is given by\cite{TKNg}:
\begin{eqnarray}
H &=& H_0 +H_K\nonumber \\
H_K&=& -a \sqrt{\frac{2\pi}{K}} \frac{J_1^z}{\pi a} S_z \partial_x\theta_s(0)\nonumber \\ 
&+& \frac{2 J_2^z}{\pi a}  S_z \sin(\sqrt{\frac{2\pi}{K_\rho}}\theta_a(0)) 
\sin(\sqrt{\frac{2\pi}{K_\sigma}}\phi_a(0)) \nonumber \\
&+& \frac{J_1^{xy}}{\pi a} [S^{-} \exp^{-{\it i}(\sqrt{\frac{2\pi}{K_\sigma}}\phi_s(0))} +h.c.] 
\cos(\sqrt{\frac{2\pi}{K_\sigma}}\phi_a(0))
\nonumber \\
&+& \frac{J_2^{xy}}{\pi a} [S^{-} \exp^{-{\it i}(\sqrt{\frac{2\pi}{K_\sigma}}\phi_s(0))} +h.c.]
\cos(\sqrt{\frac{2\pi}{K_\rho}}\theta_a(0))], \nonumber \\
H_0 &=& \frac{v_F^{'}}{2} \int dx (\partial_x \phi_s)^2 + 
(\partial_x \theta_s)^2 + (\partial_x \phi_a)^2 + (\partial_x \theta_a)^2\nonumber \\ 
\label{H_weak_bosonized}
\end{eqnarray}
with 
$K_\rho = 1/K_\sigma= K= \sqrt{\frac{1+\frac{g_4}{2}\pi v_F-
\frac{g_2}{2}\pi v_F}{1+\frac{g_4}{2}\pi v_F+\frac{g_2}{2}\pi v_F}}$ 
being the Luttinger parameter, $v_F^{'} = v_F \sqrt{(1+\frac{g_4}{2}\pi v_F)^2 
-(\frac{g_2}{2}\pi v_F)^2}$. 
Here, we consider repulsive electron-electron 
interactions ($g_2, g_4 >0$), giving $0<K<1$; and 
$K_{\rho (\sigma)}$ refers to the interaction strength in the 
charge (spin) sector. Note that $K_\sigma = 1, K_\rho <1$ 
corresponds to a spinful Luttinger liquid with 
SU(2) spin symmetry; while as $K_\sigma \neq 1$ when this symmetry 
is broken. The helical 
Luttinger leads in $H_0+H_{int}$ we consider here corresponds to a spinful 
Luttinger liquid lead with broken SU(2) symmetry and $K_\sigma>1$\cite{Kane}. 
Note that we have dropped the Klein factors in Eq.~\ref{H_weak_bosonized} 
as they can be included straightforwardly in the same manner 
as shown in Refs.~\cite{bosonizedRG,florens}.
\begin{figure}[t]
\begin{center}
\includegraphics[width=8.5cm]{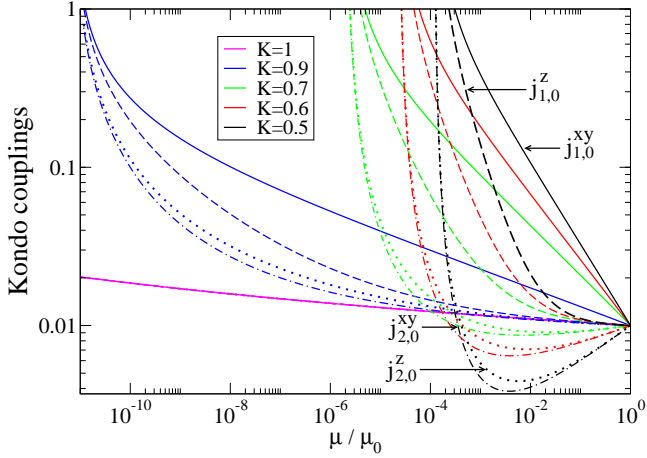}
\end{center}
\par
 \vskip -0.7cm
\caption{
(Color online)
RG flows of various Kondo couplings in the weak-coupling regime with fixed 
bare Kondo couplings: $j_{i,0}^{xy,z} = 0.01 \mu_0$ and $\mu_0=1$. Curves with different colors indicate the RG flows with different Luttinger parameters $K$. The solid, dashed, dotted, and dot-dashed lines represent the RG flows for 
$j_{1}^{xy}$, $j_{1}^z$, $j_{2}^{xy}$, and $j_2^z$, respectively.
}
\label{RG-weak}
\end{figure}

\section{ RG analysis of the model.}
\subsection{\bf RG analysis in weak coupling fixed point: $J_i^{xy,z} =0$.}
\subsubsection{RG scaling equations.}
In the vicinity of the fixed point $J_i^{xy,z} =0$, 
it has been shown in Ref. \cite{TKNg} 
that the scaling dimensions of these Kondo 
couplings based on the bosonized Hamiltonian Eq. 1 of Ref. \cite{TKNg} 
are: $[J_1^{xy}]=K <1, [J_1^{z}]=1, [J_2^{xy}]=[J_2^z]= \frac{1}{2}(K+\frac{1}{K})>1$ 
where $K <1$ for repulsive electron interactions. 
The authors of Ref. \cite{TKNg} 
argued that for $K<1$, under renormalization group (RG) transformations, 
the 2CK fixed point is reached as  the relevant $J_1$ couplings flow 
to large values with decreasing temperatures; 
while the irrelevant $J_2$ couplings decrease to zero. However, 
for $K$ being slightly less than $1$, $K\rightarrow 1^-$, via 
1-loop RG, beyond the bare scaling dimension analysis, 
we find it is possible that all four Kondo couplings can   
flow to large values, depending on the values of $K$ and the values of 
the bare Kondo couplings. This seems to suggest the possible existence of the  
1CK fixed point in the parameter space of the model.

Following Refs.~\cite{bosonizedRG,bosonization} via the renormalization group 
analysis of the bosonized Kondo model Eq.~\ref{H_weak_bosonized}, the 
1-loop RG scaling equations in the limit of $K\rightarrow 1^{-}$ read 
(see Appendix A.):
\begin{figure}[t]
\begin{center}
\includegraphics[width=8.5cm]{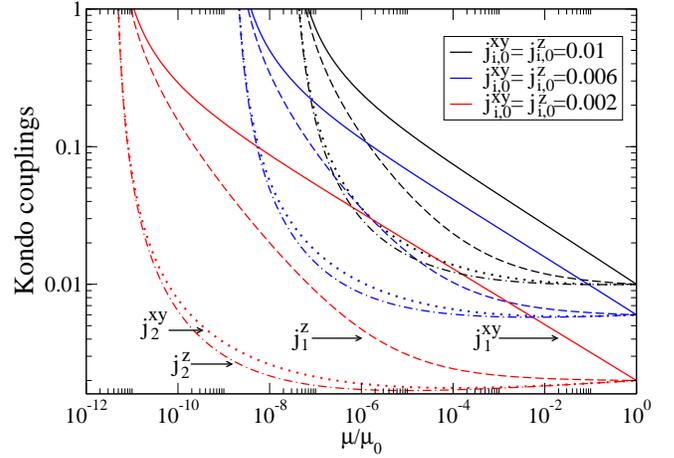}
\end{center}
\par
 \vskip -0.7cm
\caption{
(Color online)
RG flows of various Kondo couplings in the weak-coupling regime with fixed 
$K=0.8$ for various bare Kondo couplings (in units of $\mu_0=1$).  
The solid, dashed, dotted, and dot-dashed lines represent the RG flows for 
$j_{1}^{xy}$, $j_{1}^z$, $j_2^{xy}$, and $j_{2}^{z}$, respectively.
}
\label{RG-fixed-K}
\end{figure}
\begin{eqnarray}
\frac{\partial j_{1}^{xy}}{\partial \ln \mu} &=& (K-1)j_1^{xy} - j_1^{xy} j_1^z -  
j_2^{xy} j_2^z, \nonumber \\
\frac{\partial j_{1}^{z}}{\partial \ln \mu} &=& - (j_2^{xy})^2 - (j_1^{xy})^2, \nonumber \\
\frac{\partial j_{2}^{xy}}{\partial \ln \mu} &=& 
[\frac{1}{2}(K+\frac{1}{K})-1] j_2^{xy}- j_2^{xy} j_1^z -  j_1^{xy} j_2^z, \nonumber \\
\frac{\partial j_{2}^{z}}{\partial \ln \mu} &=& 
[\frac{1}{2}(K+\frac{1}{K})-1] j_2^{z}- 
2 j_1^{xy}j_2^{xy} \nonumber \\
\label{RGweak}
\end{eqnarray}
where $\mu$ is the running cutoff energy scale, and 
the dimensionless Kondo couplings are defined as: 
$j_1^{xy} \equiv \rho_0 \mu^{K-1} J_1^{xy}$, $j_1^{z} = \rho_0 J_1^{z}$, 
$j_2^{xy} = \rho_0  \mu^{\frac{1}{2} (K+\frac{1}{K})-1} J_2^{xy}$, 
and $j_2^{z} = 
\rho_0 \mu^{\frac{1}{2}(K+\frac{1}{K})-1} J_2^{z}$ with $\mu$ 
being a cutoff energy scale, 
and $\rho_0 = \frac{1}{\pi v_F^\prime}\equiv \frac{1}{2\mu_0}$ being the 
constant density of states for non-interacting leads ($K=1$) and 
$\mu_0=1$ being the band width of the conduction electrons in the leads. 
Note that the linear term in the above RG scaling equations comes from 
the non-trivial scaling dimensions of the corresponding Kondo couplings; while 
the quadratic terms in Kondo couplings are the corrections at 1-loop order. 

For $K \rightarrow 1^-$, 
both $J_2^{xy,z}$ terms are marginally irrelevant, 
$[J_2^{xy,z}] \rightarrow 1^-$. Therefore, 
within the validity of perturbative RG, both $J_2^{xy,z}$ terms can still 
flow to a large value  if bare Kondo couplings $J_{i}^{xy,z}$ 
are large enough (but they are still small, 
$J_i^{xy,z}= \mathcal{O}(1-K) \ll 1$) or the 
electron interactions in the leads are weak enough, 
leading to (possibly) 1CK fixed point ({\it ie.}, the quadratic terms 
overcome the linear term in RG equations). However, for small enough 
bare Kondo couplings (or strong enough interactions in the leads, $K\ll 1$), 
$J_2^{xy,z}$ terms are irrelevant and hence 
the system moves towards the 2CK fixed point. 
As shown in Fig.~\ref{RG-weak}, in the 
relatively higher temperature (energy) regime $10^{-3}<\mu/\mu_0<1$, with decreasing $K$ the system tends to flow to 2CK fixed point where $J_1^{xy,z}$ 
flow to large values while $J_2^{xy,z}$ 
decreases with decreasing temperature (energy). 
On the other hand, for weak enough interactions in the leads, 
$K\rightarrow 1^{-}$, all four Kondo couplings tend to flow to 1CK fixed 
point with large values (see Fig.~\ref{RG-weak}). 
Similar trend is found for a fixed $K\rightarrow 1$ 
and different bare Kondo couplings as shown in Fig.~\ref{RG-fixed-K}.
It is therefore reasonable to expect a 
1CK-2CK quantum phase transition in the parameter space 
of ${J_{1,2}^{xy,z}, K}$. However, the weak coupling RG analysis 
is valid only at relatively higher energies, and it breaks down 
as the system gets closer to the ground state, which explains 
the rapid increase of $J_2^{xy,z}$ in Fig.~\ref{RG-weak} and 
Fig.~\ref{RG-fixed-K} at lower temperatures where $J_1^{xy}$ already 
exceeds the perturbative regime, $J_1^{xy}> 1$. 
In fact, the low energy behaviors  
are determined by the physics in the strong coupling regime. 
Therefore, to address the possible quantum phase transition between 
1CK and 2CK 
fixed points, it is necessary to be able to access the neighborhood of 
the strong-coupling 
2CK fixed point as we shall discuss below.

\subsubsection{2-channel Kondo temperature $T_K^{2CK}$.}
\begin{figure}[t]
\begin{center}
\includegraphics[width=8.5cm]{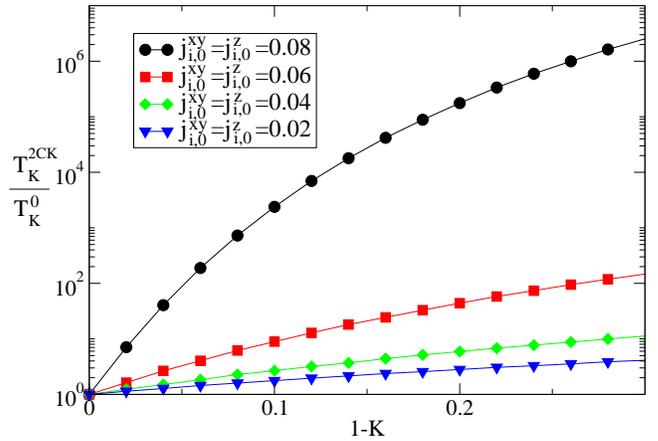}
\end{center}
\par
 \vskip -0.7cm
\caption{
(Color online)
The 2CK Kondo temperature $T_K^{2CK}$ (in units of $T_K^0$, the Kondo 
temperature of the corresponding non-interacting leads, $K=1$) 
as a function of $1-K$ for various bare Kondo couplings (in units of $\mu_0=1$) 
via the weak-coupling RG analysis. 
}
\label{weak_Tk_K}
\end{figure}

To probe the crossover between 1CK and 2CK fixed points, 
it is instructive to investigate how the Kondo temperature $T_K$ changes with 
increasing electron-electron interaction in the leads (or with 
decreasing the value of $K$ from $1$). Since $J_1^{xy}$ becomes more 
relevant with decreasing $K$ in the weak-coupling regime ($[J_1^{xy}]= K<1$), 
it is expected that under RG the system first flows very quickly 
to the vicinity of 2CK fixed point. 
As shown in Fig.~\ref{weak_Tk_K}, we find the Kondo temperature 
$T_K^{2CK}$ associated with the 2CK fixed point, defined as the energy scale 
$\mu =T_K^{2CK}$ under RG where $J_1^{xy}, J_1^{z}\approx \mathcal{O}(1)$, 
increases rapidly with increasing electron interactions 
in the leads, and its value is much larger than the Kondo temperature 
of the same setup in the non-interacting limit ($K=1$) $T_K^0$, $T_K^{2CK}\gg T_K^0$. By contrast, in the case of a Kondo dot coupled to ordinary spinful 
Luttinger liquid leads 
in Refs.~\cite{kim,gogolin}, $J_1^{xy}$ is a marginal operator 
at tree-level ($[J_1^{xy}]=1$) in the weak-coupling limit; therefore,  
the 2CK energy scale $T_K^{2CK}$ is 
much smaller than the Kondo scale for the corresponding 
non-interacting leads $T_K^0$, $T_K^{2CK}\ll T_K^0$. 
Though the system in the weak coupling regime 
quickly approaches the strong-coupling 2CK fixed point 
as $\mu \rightarrow T_K^{2CK}$, 
the ultimate fate of the ground state depends 
on the RG flows of various Kondo couplings in the strong coupling regime 
as discussed below. 

\subsection{\bf RG analysis near strong coupling  (2CK) fixed point.}
\subsubsection{\bf RG scaling equations and the phase (RG flow) diagram}
The authors in Ref. \cite{TKNg} performed scaling dimension analysis near 
a strong coupling regime where $J_1^{z} =\mathcal{O}(1), J_2^{xy,z}= J_1^{xy}=0$. They performed the Emery-Kivelson unitary transformation\cite{kivelson} 
$U= e^{{\it i}\sqrt{2\pi}\phi_s}$ to the 
bosonized Hamiltonian (Eq. 1 of Ref.\cite{TKNg}), 
and arrived Eq. 2 of Ref. \cite{TKNg}. 
\begin{eqnarray}
H &=& H_0 - \tilde{J}_1^z S_z \partial_x\theta_s(0)\nonumber \\ 
&+& \frac{2 J_2^z}{\pi a} S_z \sin(\sqrt{\frac{2\pi}{K}}\theta_a(0)) 
\sin(\sqrt{2\pi K}\phi_a(0)) \nonumber \\
&+& (S^{-}+S^{+}) [\frac{J_1^{xy}}{\pi a} 
\cos(\sqrt{2\pi K}\phi_a(0)) \nonumber \\
&+& \frac{J_2^{xy}}{\pi a} 
\cos(\sqrt{\frac{2\pi}{K}}\theta_a(0))] \nonumber \\
\label{H_K-2CK}
\end{eqnarray}
with $-\tilde{J}_1^z = \sqrt{2\pi K} v_F -\sqrt{\frac{2}{\pi K}} J_1^z $. 

They found the scaling dimensions for the Kondo couplings to be 
$[J_1^{xy}]=\frac{K}{2}$, $[J_2^{xy}]= \frac{1}{2K}$, $[J_2^z]= 
\frac{1}{2}(K+\frac{1}{K})$. 
The $J_1^{xy}$ term is relevant for $K<1$, 
and $J_2^{xy}$ term is relevant for $\frac{1}{2}<K<1$. As $K\rightarrow 1^{-}$, 
$J_2^z$ becomes marginally irrelevant, and it can flow to a large 
value under RG if the bare Kondo couplings are large enough once the 1-loop 
RG is performed. 
This seems to suggest a stable 1CK near strong coupling regime 
as all of the four Kondo couplings can either flow to or stay at large values 
(of order $1$).

To gain more insight into the stability of the 1CK/2CK fixed point, 
we apply RG approach at 1-loop order together with bosonization 
and re-fermionization near 2CK fixed point. 
First, we shall map the bosonized Hamiltonian Eq. 1 of Ref. \cite{TKNg} 
onto an effective Kondo model via re-fermionization. It has been shown 
in Ref.~\cite{TKNg} that near the strong coupling 2CK 
fixed point $J_1^{xy}\rightarrow \infty, J_1^z \rightarrow \mathcal{O}(1), 
J_2^{z/xy}\rightarrow 0$, the effective Hamiltonian reads: 
$H_{2CK}= H_0 + \frac{2J_1^{xy}}{\pi a} S^x \cos(\sqrt{2\pi K}\phi_a(0))$.  
Note that near 2CK fixed point, the dominating ``backscattering'' 
$J_1^{xy}$ term effectively cuts the Luttinger wire into two separate 
pieces at $x=0$\cite{TKNg,bosonization}, 
leading to the well-known open boundary condition 
for an impurity in a Luttinger liquid at $x=0$: $c_{i,R}(0) = -c_{i,L}(0)$ 
(or $\phi_{iR}(0) = -\phi_{iL}(0)$). The boson field 
$\phi_a$ is approximately 
pinned to a constant value\cite{TKNg}. 
Also, since $S_x$ commutes with $H_{2CK}$, we may 
therefore set $S_x$ to its eigenvalue $\pm \frac{1}{2}$ in $H_{2CK}$.
The the scaling dimensions of 
Kondo couplings near 2CK fixed point 
are\cite{TKNg,kane-fisher,bosonization} 
$[J_2^{xy}] = \frac{1}{K}, [\tilde{J}_1^z]= 1+ \frac{1}{2K}, [J_2^z] 
= \frac{1}{K} + \frac{K}{2}$. Note that all the above three couplings 
are irrelevant for $K<1$. This suggests that the system favors the 
2CK fixed point at ground state $K<1$. Meanwhile, by a stability analysis 
in Sec. V., we show that the  
2CK fixed point is a also a stable fixed point for $K>\frac{1}{2}$ 
once the system gets there. 

However, as suggested in our weak-coupling RG analysis, 
the 2CK fixed point may be unstable for $K\rightarrow 1^{-}$ and/or  
large enough bare Kondo couplings such that $J_2^{xy,z}$ may 
become relevant again, and the system can  
undergo a 1CK-2CK quantum phase transition. 
To address this possibility, we shall focus below on the 1-loop 
RG flows of the leading two irrelevant operators near 
the 2CK fixed point, given by: 
\begin{equation}
\delta H_{2CK}= \frac{J_2^{xy}}{\pi a} S^x \cos(\sqrt{\frac{2\pi}{K}} \theta_a(0)) 
- \tilde{J}_1^z S_z \partial_x \theta_s(0).
\label{delta-H-2CK}
\end{equation}
Via the similar re-fermionization as shown in Appendix A., 
we map $H_0 + \delta H_{2CK}$ onto an effective 
Kondo model subject to a bosonic environment:
\begin{eqnarray}
H_0 + \delta H_{2CK} &\rightarrow& H_0 + H_0^\prime + \delta H_{2CK}\nonumber \\
&=& \tilde{H}_0 + H_b+ \tilde{H}_{2CK},\nonumber  \\
H_0^\prime &=&  \frac{v^{'}_F}{2} \int dx 
(\partial_x \theta_{a}^\prime)^2\nonumber \\
 \tilde{H}_0 &=& \frac{v^{'}_F}{2} \int dx [2 (\partial_x \theta_{0,a})^2 
+ (\partial_x \theta_s)^2\nonumber \\
 &+&  (\partial_x \phi_s)^2 + (\partial_x \phi_a)^2]\nonumber \\
&=& \sum_{k,\sigma,i=1(\tilde{L}),2(\tilde{R})} \epsilon(k) 
\tilde{c}^{\dagger\sigma}_{k,i} 
\tilde{c}_{k,i}^{\sigma}, \nonumber \\
H_b &=& \frac{v^{'}_F}{2} \int dx 
2 (\partial_x \tilde{\theta}_a)^2, \nonumber \\ 
\tilde{H}_{2CK} &=& J_2^{xy} S^- [
s_{\tilde{L}\tilde{R}}^+ e^{{\it i} 
\sqrt{4\pi (\frac{1}{K}-1)}\tilde{\theta}_a(0)}\nonumber \\ 
&+& s_{\tilde{R}\tilde{L}}^+ 
e^{-{\it i} \sqrt{4\pi (\frac{1}{K}-1)} \tilde{\theta}_a(0)}] 
+  h.c. \nonumber \\ 
&+& \sqrt{\frac{\pi}{2}}\tilde{J}_1^z (s_{\tilde{L}\tilde{L}}^z + s_{\tilde{R}\tilde{R}}^z) S_z 
\label{dH2CK}
\end{eqnarray}
where the boson field $\theta_a^\prime$ in 
$H_0^\prime$ is decoupled from $H_0$ and is added here just 
for the mapping, 
the effective non-interacting electron operator 
$\tilde{c}_{k,i}^{\sigma}$ is defined in Eq.~\ref{tilde-c}. 
Note that  
since the scaling dimension of 
$\cos(\sqrt{\frac{2\pi}{K}}\theta_a)$ at 2CK 
fixed point in Eq.~\ref{dH2CK} is $\frac{1}{K}$ due to open boundary 
condition\cite{TKNg}, 
we have made the following decomposition for the boson field 
$\sqrt{\frac{1}{K}}\theta_a$:
\begin{eqnarray}
\sqrt{\frac{1}{K}} \theta_{a} &=& 
\sqrt{2}\theta_{0,a} +  \sqrt{2} \bar{\theta}_a,\nonumber \\
\sqrt{\frac{1}{K}} \theta_{a}^\prime &=& 
\sqrt{2 (\frac{1}{K}-1)}\theta_{0,a} - 
\sqrt{\frac{2}{\frac{1}{K}-1}} \bar{\theta}_a,\nonumber \\
  \theta_{s} &=& \theta_{0,s},\nonumber \\
\bar\theta_a &=& \sqrt{\frac{1}{K}-1} \tilde{\theta}_a.  
\label{2CK-decompose}
\end{eqnarray}
The re-fermionization of $H_0$ is done through the following identifications:
\begin{eqnarray}
\sqrt{2} \theta_{0,a} &=& \sqrt{2}(\phi_{0,1}^{\uparrow} + 
\phi_{0,2}^{\downarrow}) 
=  -\sqrt{2}(\phi_{0,2}^{\uparrow} + \phi_{0,1}^{\downarrow}),\nonumber \\
\sqrt{2} \theta_{0,s} &=&  \phi_{0,1}^{\uparrow} - \phi_{0,1}^{\downarrow} +   
\phi_{0,2}^{\uparrow} - \phi_{0,2}^{\downarrow} \nonumber \\
\label{id-2CK}
\end{eqnarray}  
where we have decomposed the boson field 
$\sqrt{\frac{1}{K}}\theta_{a}$ into two independent sets of 
boson fields: the ``free'' 
($\theta_{0,a}$) and ``interacting'' 
($\tilde{\theta}_{a}$) parts:
Here, the ``free'' part of the boson fields 
$\theta_{0,a}$ (defined in the same way as 
in Sec.II.) can be re-fermionized 
into two effective non-interacting 
fermion leads described by $\tilde{H}_0$ with 
$\tilde{c}_{\alpha}^{\sigma}$ being the electron 
destruction operator of the effective non-interacting leads
 \begin{equation}
\tilde{c}_{i=1(\tilde{L}),2(\tilde{R})}^{\uparrow (\downarrow)} 
= \frac{1}{\sqrt{2\pi a}} 
F_{i}^{\uparrow (\downarrow)} 
e^{\pm {\it i}(\sqrt{4\pi} \phi_{0,i}^{\uparrow (\downarrow)}(x) + k_F x)}.
\label{tilde-c}
\end{equation}
with $\alpha=1(\tilde{L}), 2(\tilde{R})$ being the index for effective 
non-interacting leads, $s_{\gamma \beta }^{\pm(z) }=\sum_{\alpha ,\delta ,k,k^{\prime }}\frac{1}{2}\tilde{c}_{k\gamma}^{\dagger \alpha}\mathbf{\sigma }_{\alpha \delta }^{\pm(z) }\tilde{c}_{k^{\prime
}\beta}^{\delta}$ being the spin-flip (z-component of the spin) 
operators between the effective leads $\gamma $ and $\beta$. 
Note that the effective non-interacting leads also exhibit
the helical nature; namely, the spin up/down ($\sigma=\uparrow/\downarrow$)
electrons are tied to the right (R)/left (L) moving particles, respectively. 
The ``free'' part of boson field $\theta_{0,a}$ 
follow the correlations of the free fremions 
in 1D:
\begin{eqnarray}
<e^{-{\it i} \sqrt{2\pi} \theta_{0,a}(t)} 
e^{{\it i} \sqrt{2\pi} \theta_{0,a}(0)}> &\propto& 
\frac{1}{t}. \nonumber \\
\label{theta-a}
\end{eqnarray}
Meanwhile, $H_b$ represents for the 
effective dissipative ohmic boson environment (baths) 
made of the ``interacting'' part of the effective bosons 
$\tilde\theta_{a}$. 
These bosons couple to the Kondo dot through the 
additional exponential ``phase'' factors in the 
effective Kondo terms $\tilde{H}_{2CK}$, leading 
to all the combined Kondo-Luttinger physics\cite{florens}. 
In particular, since 
these dissipative ohmic bosons obey the following 
correlations  via Eq.~\ref{dH2CK}: 
\begin{equation}
< e^{-{\it i} \sqrt{4\pi (\frac{1}{K}-1)}\tilde\theta_a (t)}  
 e^{{\it i} \sqrt{4\pi (\frac{1}{K}-1)}\tilde\theta_a (0)}> 
\propto \frac{1}{t^{2(\frac{1}{K}-1)}};
\end{equation} 
while  the impurity spin 
operator $S_z$ exhibits the following correlation\cite{TKNg}: 
\begin{equation}
<S_z(0) S_z(t)>\propto 
\frac{1}{t^{\frac{1}{K}}}. 
\end{equation}
These correlations lead to the non-trivial bare scaling dimensions 
of the Kondo couplings 
and therefore to the first term (linear in the Kondo coupling) 
of the RG scaling equations. Note that since near 2CK fixed point 
$\phi_{s,a}$ fields are decoupled from Eq.~\ref{delta-H-2CK}, we have 
effectively two independent degrees of freedom left among the four: 
$(\phi_{s,a},\theta_{s,a})$; the open boundary condition for 
the spin-up right-moving (R) and spin-down left-moving electrons:  
$\phi_{0,i}^{\uparrow (R)} = 
-\phi_{0,i}^{\downarrow (L)}$ is implied in Eq.~\ref{id-2CK}. 
With the help of Eq.~\ref{2CK-decompose} 
and Eq.~\ref{id-2CK},         , 
we finally arrive $\tilde{H}_0$ and $\tilde{H}_{2CK}$ in Eq.~\ref{dH2CK}.

Next, we shall obtain the one-loop RG scaling equations for $J_2^{xy}$ and 
$\tilde{J}_1^z$ in Eq.\ref{dH2CK}. To this aim, we define 
the dimensionless couplings 
$j_2^{xy}\equiv \rho_0\tilde{c}_2^{\perp} 
\mu^{\epsilon} J_2^{xy}(\mu)$ and $j_1^z \equiv 
\rho_0 \tilde{c}_1^z \mu^{\epsilon^{\prime}} \tilde{J}_1^z(\mu)$ 
and  $\epsilon\equiv \frac{1}{K}-1$, 
$\epsilon^{\prime}\equiv \frac{1}{2K}$ with $\tilde{c}_1^z$, $\tilde{c}_2^{\perp}$ being defined in Appendix B. and C.. 

We derive the 1-loop RG scaling equations via the poor-man's scaling 
approach in Ref.~\cite{florens} (see Appendix B.) and via field-theoretical 
$\epsilon$-expansion technique (see Appendix C.): 
\begin{eqnarray}
\frac{\partial j_{2}^{xy}}{\partial \ln \mu} &=& 
\epsilon j_{2}^{xy} -j_{2}^{xy} j_{1}^z,\nonumber \\
\frac{\partial j_{1}^{z}}{\partial \ln \mu} &=& \epsilon^{\prime} j_{1}^{z} -(j_{2}^{xy})^2.
\label{RG2CK}
\end{eqnarray}

\begin{figure}[t]
\begin{center}
\includegraphics[width=6cm]{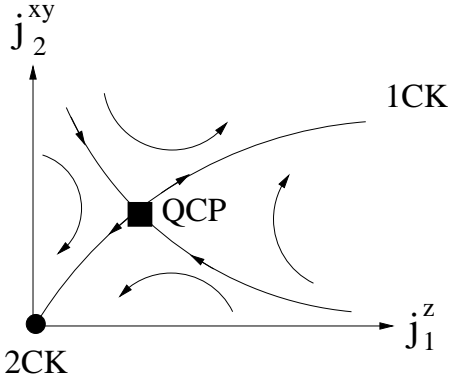}
\end{center}
\par
 \vskip -0.7cm
\caption{
(Color online)
Schematic diagram of the RG flow near 2CK fixed point. The 1CK-2CK 
quantum critical point (QCP) is represented by the filled 
black square located at 
$j_c=(j_{1c}^z, j_{2c}^{xy}) = (\epsilon, \sqrt{\epsilon\epsilon^{\prime}})$. }
\label{phasedia}
\end{figure}
\begin{figure}[t]
\begin{center}
\includegraphics[width=9cm]{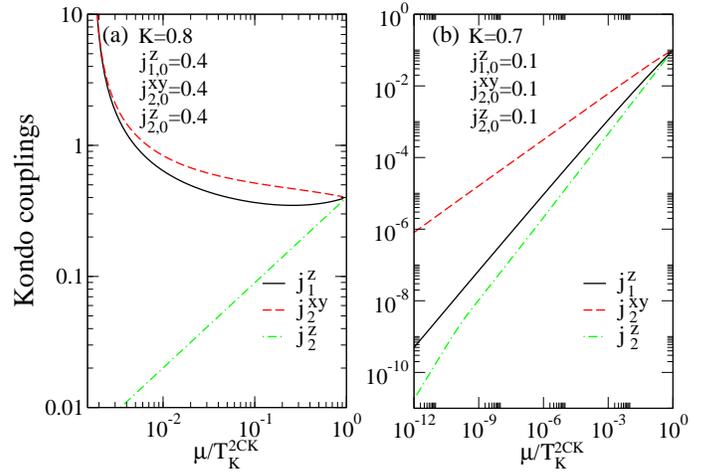}
\end{center}
\par
 \vskip -0.7cm
\caption{
(Color online)
RG flows of various Kondo couplings in the strong-coupling 
regime near 2CK fixed point with the following parameters: 
(a). $j_{i,0}^{xy,z} (\mu=T_K^{2CK})= 0.4 \mu_0$, 
$K=0.8$ and (b). $j_{i,0}^{xy,z} (\mu=T_K^{2CK}) = 0.1 \mu_0$, $K=0.7$. 
The RG flows starting from $\mu=T_K^{2CK}$ to $\mu\rightarrow 0$ 
in (a) are towards the 1CK fixed point; while as the flows in (b) 
are moving towards the 2CK fixed point. 
The solid, dashed, and dot-dashed lines represent the RG flows for 
$j_{1}^{z}$, $j_{2}^{xy}$, and $j_{2}^{z}$, respectively. Here, we set $\mu_0=1$.
}
\label{2CK-RGflow}
\end{figure}

For $K\rightarrow 1^-$ we find an intermediate quantum critical 
fixed point (QCP) 
at $j_c=(j_{1c}^z, j_{2c}^{xy})=(\epsilon, \sqrt{\epsilon\epsilon^{\prime}})$ 
within the validity 
of the perturbative RG separating the 1CK fixed point (for 
$j_0 \equiv (j_{1,0}^{z}, j_{2,0}^{xy}) > j_c \equiv (j_{1,c}^{z}, j_{2,c}^{xy})$) 
and 2CK fixed point (for $j_0<j_c$) 
where $j_2^{xy}$ and $j_1^z$ flow towards a large and 
vanishingly small value, respectively 
(see Fig.~\ref{phasedia}). The RG flows near 
the QCP are determined by linearizing the RG 
scaling equations as shown in Fig.~\ref{phasedia}. The typical 
RG flows corresponding to the 1CK and 2CK fixed points are shown 
in Fig.~\ref{2CK-RGflow} (a) and (b), respectively.  

Note that $j_1^z\rightarrow 0$ near 2CK fixed point suggests that 
the original Kondo coupling $J_1^z$ (see Eq.~\ref{H_K-2CK}) 
is at a large value (order of $1$): $J_1^z\approx \mathcal{O}(1)$, 
consistent with the familiar 2CK fixed point 
with both $J_1^{xy,z}$ being large. However, we find the ``1CK'' 
fixed point here in the strong-coupling analysis 
seems somewhat different from  
the familiar (conventional) 
1CK fixed point we obtained in the weak coupling regime where 
all the four Kondo couplings will flow to (or stay at) large values.  
Instead, our RG analysis based on re-fermionization 
for the coupling 
$j_2^{z}=\rho_0 \tilde{c}_2^z \mu^{\frac{1}{K}+\frac{2}{K}-1} J_2^z$ 
in Eq.~\ref{H_K-2CK} near 2CK fixed point 
shows that it stays irrelevant up to 1-loop order with the RG 
scaling equation:    
\begin{equation}
\frac{\partial j_{2}^{z}}{\partial \ln \mu} =
\left[\frac{1}{K}+ \frac{K}{2}-1\right]j_{2}^{z}  
\label{j2z}
\end{equation} 
where we find no corrections at 1-loop order. 
Note that unlike in the weak coupling RG where $j_1^{xy} j_2^{xy}$  
will contribute to the 1-loop renormalization of $j_2^z$ (see Eq.\ref{RGweak}, 
the $j_1^{xy} j_2^{xy}$ term is absent here in Eq.~\ref{j2z} as $j_1^{xy}$ is 
already very large near the 2CK fixed point, 
$j_1^{xy}(T\approx T_K^{2CK})\gg 1$.

It is clear from Eq.~\ref{j2z} that\cite{TKNg} 
$j_2^{xy}(\mu) \propto 
\mu^{\frac{1}{K}+ \frac{K}{2}-1}$, 
vanishing as $\mu\ll T_K^{2CK}$ even for $j_0>j_c$ 
where the system eventually 
flows to the 1CK fixed point (see Fig.~\ref{2CK-RGflow}). 
By combining the 1-loop RG analysis in the weak and 
strong coupling limits, we may obtain the full crossover of 
$J_2^z$ for the system which will eventually flow to the 1CK fixed point: 
For $T_K^{2CK}<T<\mu_0$, $j_2^z$ first grows 
to order of $1$ (see Fig.\ref{RG-weak}); 
then it vanishes in a power-law fashion 
at lower temperatures $T\ll T_K^{2CK}$ (see Fig.\ref{2CK-RGflow}).
However, the above qualitative feature for $j_2^z$ based on the 
1-loop RG analysis might get modified at the 2-loop order, 
which exceeds the scope of our current work 
and will be addressed elsewhere. 

Though somewhat unconventional, the fixed 
point with $j_{1,2}^{xy},j_1^z\rightarrow \infty$ and $j_2^z\rightarrow 0$ 
can still be regarded as the 
one-channel Kondo (1CK) fixed point since the two leads are connected by the 
strong transverse Kondo couplings $j_2^{xy}$; and only 
one channel of conduction electrons (even combination of the two leads) 
couples to the Kondo dot. Therefore, we expect the 
linear conductance $G_{\perp}(T)$ contributed from $J_2^{xy}$   
at the 1CK fixed point here to show the same temperature dependence 
as those in the isotropic one-channel Kondo system.

\subsubsection{1-channel Kondo temperature $T_K^{1CK}$}

\begin{figure}[t]
\begin{center}
\includegraphics[width=8.5cm]{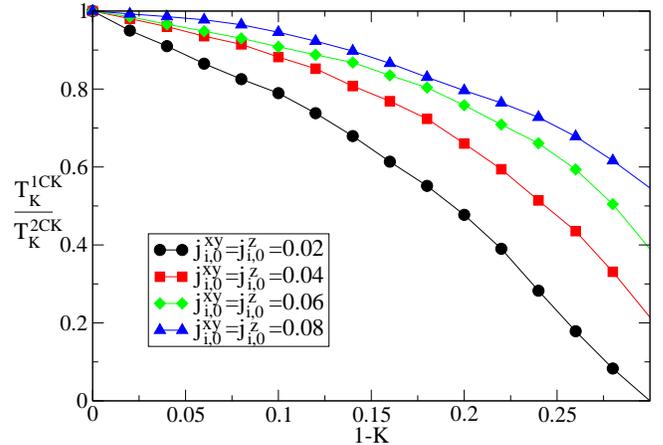}
\end{center}
\par
 \vskip -0.7cm
\caption{
(Color online)
The ratio of 1CK to 2CK Kondo temperature $T_K^{1CK}/T_K^{2CK}$ 
as a function of $1-K$ for various bare Kondo couplings 
$j_{i,0}^{xy,z}(\mu=\mu_0)$ (in units of $\mu_0=1$) 
via the weak-coupling RG analysis. 
}
\label{strong_Tk_K}
\end{figure}

As mentioned above, for $j_0>j_c$ with decreasing temperature 
the system crosses over from 2CK 
to 1CK fixed point at a much lower energy scale 
$\mu \approx T_K^{1CK} \ll T_K^{2CK}$ where $T_K^{1CK}$ refers to the 
Kondo temperature associated with the 1CK fixed point. 
As shown in Fig.~\ref{strong_Tk_K}, the 1CK fixed point persists to be the 
ground state at a finite but weak electron-electron interaction strength, 
$K_c<K<1$ with $K_c$ being the critical interaction below which 
the ground state switches from 1CK to 2CK fixed point. Meanwhile, the crossover 
scale to 1CK fixed point $T_K^{1CK}$ (with respect to $T_K^{2CK}$) 
for $j_0>j_c$ gets reduced significantly as interaction gets stronger. Also, 
for a fixed value of $K$, the ratio $T_K^{1CK}/T_K^{2CK}$ is larger 
for larger bare Kondo couplings $j_0$, as expected. 

\section{Stability analysis of 1CK and 2CK fixed point for $K<1$.}
Having found the possible QPT between 1CK and 2CK fixed points, 
it is important to perform a stability analysis and study 
how robust the quantum critical point 
of our system is against small perturbations. Equivalently, we 
need to know how stable the 1CK and 2CK fixed points are for $K<1$. 

We first examine the stability of the helical Luttinger liquid lead itself. 
In general there exists the 
single particle backscattering term due to the interaction 
of $c_{i,R/L}(0)$ and the quantum dot\cite{TKNg}:  
$t' c^{\dagger}_{i,R}c_{i,L} + h.c.$. However, this term 
is forbidden here as it breaks time-reversal symmetry. 
Meanwhile, for 1-D Hubbard model in general there 
exists the ``spin-flip'' backscattering term in $H_u$ of the form: 
$H_{sf}\propto c_{i,L}^{\dagger\uparrow}c_{i,L}^{\downarrow}
c_{i,R}^{\dagger\downarrow}c_{i,R}^{\uparrow} + h.c.$. However, 
due to the helical nature of our leads (or the Right/Left moving electrons 
are tied to their spins, {\it i.e.}, only 
$c_{R(L)}^{\uparrow(\downarrow)}$ electrons exist), 
this $H_{sf}$ term is therefore absent. Nevertheless, the 
the Umklapp term that exists on a single bond is allowed by the 
time-reversal symmetry\cite{SCZhang}: 
\begin{eqnarray}
H_{um} &=& g_u c^{\dagger\uparrow}_R(0) 
c^{\dagger\uparrow}_R(0) c_L^{\downarrow}(0) \nonumber \\
&\times& c_{L}^{\downarrow}(0) + h.c.\nonumber \\
\end{eqnarray}
The scaling dimension of this term has 
been shown to be $[H_{um}]=4K$, suggesting that the 
helical edge state is unstable towards an insulating 
phase for $K<\frac{1}{4}$.

We now focus on the effects of the particle-hole (p-h) asymmetry 
on the stability of these two fixed points as indicated 
in Refs.~\cite{gogolin,florens,kim} that it is the most relevant 
perturbation for a Kondo dot 
coupled to Luttinger liquid leads. 
Let us first address this issue at 2CK fixed point where 
the two leads are effectively disconnected. 
The particle-hole asymmetry in our Kondo model 
generates potential scattering terms of the 
following form\cite{gogolin}:
\begin{eqnarray}
H_{1ps} &=& H_t + H_{te}, \nonumber \\
H_t &=& t \sum_{k,i,\sigma,\alpha=L,R} c^{\dagger\sigma\alpha}_{k,i} 
c_{k,i}^{\sigma\alpha},\nonumber \\ 
H_{t_e} &=& t_e
\sum_{k,i\neq j\sigma} c^{\dagger\sigma\alpha}_{k,i} c_{k,j}^{\sigma\alpha} 
+ h.c.\nonumber \\ 
\end{eqnarray}
with $i,j=1(\tilde{L}),2(\tilde{R})$ 
being the lead index, $\sigma =\uparrow (R), \downarrow (L)$ being 
spin index, and $R(L)$ being the right (left) moving particles. 
Here, $t$ and $t_e$ terms 
represent a chemical potential of each lead 
and a weak tunneling between the disconnected Luttinger leads\cite{Kane}. 
Meanwhile, 
two additional two-particle scattering terms $H_{2p}$ involving 
tunneling of spin ($t_{\sigma}$) and of charge ($t_{\rho}$) can be generated 
by the weak tunneling $t_e$ via 2nd-order perturbation (see Fig.2 (d) (e) (f) 
of Ref.~\cite{Kane}), given by:
\begin{eqnarray}
H_{2ps} &=& H_{t_{\sigma}} + H_{t_{\rho}}, \nonumber \\ 
H_{t_\sigma} &=& t_\sigma \sum_{k} 
c^{\dagger\uparrow R}_{k,1} c_{k,2}^{\uparrow R}
c_{k,2}^{\dagger\downarrow L} c_{k,1}^{\downarrow L}+ h.c. \\
 H_{t_\rho} &=& t_\rho \sum_{k} 
c_{k,1}^{\dagger\uparrow R} c_{k,2}^{\uparrow R} 
c_{k,1}^{\dagger\downarrow L} c_{k,2}^{\downarrow L}.\\
\label{H2p}
\end{eqnarray}  
The bosonized form of Eq.~\ref{H2p} reads\cite{Kane}: 
\begin{eqnarray}
H_{1ps} + H_{2ps} &=&  
t\frac{K}{2\pi a}  \partial_x \phi_s\nonumber \\ 
&+& \frac{t_e}{2\pi a} \cos(\sqrt{2\pi K}\phi_a(0))
\cos(\sqrt{\frac{2\pi}{K}}\theta_a(0))\nonumber \\
&+& \frac{t_\rho}{2\pi a} \cos(2\sqrt{2\pi K}\phi_a)\nonumber \\ 
&+& \frac{t_\sigma}{2\pi a} \cos(2\sqrt{\frac{2\pi}{K}}\theta_a).\nonumber \\
\end{eqnarray}
Near 2CK, $\phi_a(0)$ is a constant, therefore 
the scaling dimensions of these term gives: $[t]=1$,$[t_e]= \frac{1}{2K}$, 
$[t_\sigma]= \frac{2}{K}$ ($t_\rho\approx const.$)\cite{Kane}. 
It is clear that 
all operators are irrelevant for $\frac{1}{2}<K<2$;   
$t_e$ term becomes relevant for $K<\frac{1}{2}$, and $t_\sigma$ is relevant 
for $K>2$. 

Next, we consider the stability of the 1CK fixed point. Since the cross-channel 
Kondo coupling $J_2^{xy}$ term flows under RG 
along with $J_{1}^{xy}$ to large values 
while as $J_1^z$ stay at order of $1$, the two semi-infinite Luttinger wires 
are joined into one single infinite Luttinger wire\cite{Kim}. 
In contrast to the ``weak tunneling'' processes mentioned above 
at the 2CK fixed point, the potential scattering term generates 
the ``weak backscattering'' processes between the electrons in the 
upper and lower edges, including the single-particle backscattering term 
$v_e$, and the two-particle backscattering terms $v_\rho$, and $v_\sigma$ 
(see Fig. 2 (a) (b) (c) in Ref.~\cite{Kane}):
\begin{eqnarray}
H_{v_e} &=& v_e
\sum_{k} c^{\dagger\uparrow R}_{k,1} c_{k,2}^{\uparrow L} +  
c^{\dagger\downarrow L}_{k,1} c_{k,2}^{\downarrow R} + h.c., \nonumber \\
 H_{v_\rho} &=& v_\rho \sum_{k} 
c_{k,1}^{\dagger\uparrow R} c_{k,2}^{\uparrow L} 
c_{k,1}^{\dagger\downarrow L} c_{k,2}^{\downarrow R} + h.c.,\\
H_{v_\sigma} &=& v_\sigma c^{\dagger\uparrow R}_{k,1} c_{k,2}^{\uparrow L}
c_{k,2}^{\dagger\downarrow R} c_{k,1}^{\downarrow L}+ h.c. \\
\end{eqnarray}
In fact, there exists a duality mapping 
between the ``weak tunneling'' and ``weak backscattering'' limits\cite{Kane}:
$c_{k,2}^{\uparrow R} \rightarrow c_{k,2}^{\uparrow L}, 
c_{k,2}^{\downarrow L} \rightarrow c_{k,2}^{\downarrow R}, 
t_e \rightarrow v_e, 
t_\rho \rightarrow v_\sigma, 
t_\sigma \rightarrow t_\rho, 
K \rightarrow \frac{1}{K}$.
Note that at 1CK fixed point, $\phi_a$ is not pinned to a constant 
as opposed to that in the 2CK case. 
The scaling dimensions of these terms can be read off straightforwardly:
$[v_e] =\frac{1}{2}(K+\frac{1}{K})$, $[v_\rho]= 2K$, and 
$[v_\sigma]=\frac{2}{K}$. The $v_e$ term is always irrelevant for $K<1$, 
while the $v_\sigma$ and $v_\rho$ terms are irrelevant for $\frac{1}{2}<K<2$ 
and relevant otherwise. 

Based on the above analysis, we 
find that both 1CK and 2CK fixed point are 
stable for $\frac{1}{2}<K<1$, and 
unstable for $K<\frac{1}{2}$. 
We have checked that our analysis reproduces the well-known results 
for a Kondo dot coupled to conventional Luttinger liquid leads 
in Refs.~\cite{gogolin,kim,florens} where $[t_e]= \frac{1}{2K}$ at 2CK 
fixed point and $[v_e] = \frac{1}{2}(1+K)$ at 1CK fixed point. 

As a final remark, we consider here the parity (left-right) 
symmetric model where $J_1 = J_{LL}=J_{RR}$ with $J_{LL(RR)}$ 
being referred to the Kondo couplings involving only the left (right) lead. 
Nevertheless, parity 
asymmetry is a relevant perturbation near 2CK fixed point. 
In the presence of parity asymmetry ($J_{LL}\neq J_{RR}$), 
the system will flow to the 1CK fixed point with the large 
bare Kondo couplings\cite{florens}.

\section{Critical properties near 1CK-2CK quantum phase transition.}
The critical properties and crossovers of various thermodynamical 
quantities near this newly found 1CK-2CK QCP can be obtained via 
the above RG approach combined with the 
field-theoretical $\epsilon-$expansion technique\cite{Zinn-Justin,zarand,QMSi,kircan,lars}. 
We employ here a double-$\epsilon-$expansion with two small expansion 
parameters $\epsilon$ and $\epsilon^{\prime}$. Our approach is is valid 
for the Luttinger parameter $K\rightarrow 1^-$ as both parameters $\epsilon$ 
and $\epsilon^{\prime}$ are within perturbative regime: 
$\epsilon\rightarrow 0$, $\epsilon^{\prime}\rightarrow \frac{1}{2} <1$, 
and $\epsilon \ll \epsilon^{\prime}$. 
Following Refs. \cite{QMSi,kircan,lars}, 
we define the renormalized pseudo-fermion fields $\tilde{f}_{\sigma}$ 
and the renormalized dimensionless Kondo couplings $j$ as: 
$f_{\sigma}= \sqrt{Z_f} \tilde{f}_{\sigma}$, and 
$J_2^{xy} = \frac{\mu^{-\epsilon} Z_{j^{\perp}}}{\tilde{c}_2^{\perp} Z_f} 
j_2^{xy}$, $\tilde{J}_1^z =  
\frac{\mu^{-\epsilon^{\prime}} Z_{j^{z}}}{\tilde{c}_1^zZ_f} 
j_1^{z}$ with $Z_f$ and $Z_{j^{\perp}/z}$ being the 
renormalization factors for the impurity field and Kondo couplings, 
respectively and $\mu$ is a renormalization energy scale.  
The renormalization factors are obtained via 
minimal subtractions of poles\cite{kircan,QMSi}, given by 
(see Appendix C.): 
\begin{eqnarray}
Z_{j^{\perp}} &=& 1+ \frac{j_1^z}{\epsilon^{\prime}},\nonumber \\
Z_{j^{z}} &=& 1+ \frac{(j_2^{xy})^2/j_1^z}{2 \epsilon},\nonumber \\
Z_f &=& 1+ \frac{(j_{2}^{xy})^2}{8\epsilon}+ 
\frac{(j_{1}^{z})^2}{16\epsilon^{\prime}}.
\label{Z-factors}
\end{eqnarray} 
Within the field-theoretical RG approach, we have checked that  
the RG scaling equations in Eq.~\ref{RG2CK} can be reproduced 
via calculating the $\beta-$functions: $\beta(j_i) \equiv \mu 
\frac{\partial j_i}{\partial \mu}|_{j_{i,0}}$ with $\mu$ being an 
energy scale, $j_i = j_2^{xy}, j_1^z$ being the renormalized Kondo 
couplings and $j_{1,0} = j_{1,0}^{z}=J_1^{z}, 
j_{2,0}=j_{2,0}^{xy}=J_2^{xy}$ being the bare Kondo couplings 
(see Appendix C.). Below we discuss various 
critical properties and crossover functions based on field-theoretical 
$\epsilon$-expansion approach.\\
\subsection{Observables at criticality.}
We first calculate various observables at criticality, including correlation 
length exponent, impurity entropy, dynamical properties of the T-matrix 
and local spin susceptibility.\\ 
\subsubsection{\it Correlation length exponent $\nu$.} 
The correlation length exponent $\nu$ describes how the correlation length 
$\xi$ diverges when the system is tuned to the transition: 
$\xi\propto |t|^{-\nu}$ with $t\equiv \frac{j_0-j_{c}}{j_{c}}$ 
being the dimensionless distance to the QCP. 
It also gives the power-law vanish of the characteristic crossover energy 
scale $T^{\ast}$ close to the transition: $T^* \propto |t|^{\nu}$. 
To calculate $\nu$, we first linearize the RG scaling 
equations Eq.\ref{RG2CK} near QCP. 
The correlation length exponent $\nu$ is determined 
by the largest eigenvalue of the coupled linearized equations, found to be:
\begin{equation} 
\nu = \frac{4K}{\sqrt{1+16 K\epsilon} -1} = \frac{1}{2\epsilon}+\mathcal{O}(\epsilon^2,\epsilon^{\prime 2})  
\label{exponent-nu}
\end{equation}
where the leading order behavior $\nu \approx \frac{1}{2\epsilon}$ 
is obtained by expanding the square-root in Eq.~\ref{exponent-nu} 
in the limit of  $\epsilon \ll \epsilon^{\prime}$.

\subsubsection{\it Impurity entropy.}
The impurity contribution to the low-temperature 
entropy near QCP is obtained by a perturbative calculation of 
the impurity thermodynamic potential $\Omega_{imp}$\cite{kircan} 
with respect to the 2CK fixed point 
and taking the temperature derivative:   
$S_{imp}= \frac{\partial \Omega_{imp}}{\partial T}$. At QCP and $T=0$ 
it can be written as: 
\begin{equation}
S_{imp}^{QCP}= S_{imp}^{2CK} + 
\Delta S_{imp}.
\end{equation}
 where $S_{imp}^{2CK}= \ln \sqrt{2K} = \frac{1}{2}\ln 2K$ 
is the zero-temperature residual 
impurity entropy at 2CK fixed point which shows the existence of 
fractionally degenerate 
ground state\cite{TKNg,Affleck,Fendley},  
and $\Delta S_{imp}$ is the correction to $S_{imp}^{2CK}$ at QCP.  
Following similar renormalized perturbative calculations 
in Ref. \cite{kircan}, we find
\begin{equation}
\Delta S_{imp} = \pi^2\ln 2 [\frac{\epsilon (j_{2c}^{xy})^2}{4} 
+ \frac{\epsilon^{\prime} (j_{1c}^{z})^2}{8}]= \frac{3\epsilon^2 \pi^2 \ln 2}{32K}.
\end{equation}
Therefore, we have:
\begin{equation}
S_{imp}^{QCP}= \frac{1}{2}\ln 2K + \frac{3\epsilon^2 \pi^2 \ln 2}{32K}.
\end{equation}

\subsubsection{\it The $T-$matrix.}
The conduction electron $T-$matrix, 
$T_{\alpha\alpha'}(\omega)$, in the 
Kondo model carries important information on the scattering of the 
conduction electrons from lead $\alpha$ to lead $\alpha'$ via the impurity.  
In particular, $T_{\alpha\alpha'}(\omega)$ with $\alpha\neq \alpha'$ describes 
the transport across the dot, detectable in transport measurements.  
The $T-$matrix is determined from the conduction electron Green functions  
$G_{\alpha\alpha^{\prime}}(t)= <c_{\alpha}(0)c^{\dagger}_{\alpha^{\prime}}(t)>$ 
through $G_{\alpha\alpha^{\prime}} = 
G_{\alpha\alpha^{\prime}}^{0}\delta_{\alpha\alpha^{\prime}} + 
G_{\alpha\alpha}^{0}(\omega) 
T_{\alpha\alpha^{\prime}}(\omega)
G_{\alpha^{\prime}\alpha^{\prime}}^{0}(\omega)$\cite{paaske}. 
Near 2CK fixed point, 
$T_{\alpha\alpha^{\prime}}(\omega)$ with $\alpha\neq\alpha^{\prime}$ 
is defined through the propagator, $G_T$, of the composite 
operator $T_{\sigma\alpha}= J_2^{xy} 
e^{{\it i} \sqrt{4\pi (\frac{1}{K}-1)}\tilde\theta_a(0)}
f^{\dagger}_{\sigma}f_{\sigma^{\prime}}\tilde{c}_{\alpha}^{\sigma^{\prime}}$ (see Eq.~\ref{dH2CK}): 
$T_{\alpha\alpha^{\prime}(\omega)} = G_T(\omega)$\cite{kircan}. 

Following the similar calculations in Ref.~\cite{kircan} and Appendix C.,
 we analyze the propagator $G_T(\omega)$ near 2CK fixed point and find at 
zero temperature 
$Im(T_{\alpha\alpha^{\prime}}^{2CK}(\omega))\propto 
\frac{1}{\omega^{-\eta_{T}^{2CK}}}$ with the anomalous  
exponent at the tree level with respect to the 2CK fixed point given by 
$\eta_{T}^{2CK}= 2\epsilon$ ({\it ie.}, 
$Im(T_{\alpha\alpha^{\prime}}^{2CK}(\omega))\propto \omega^{2\epsilon}$). 
Near 1CK-2CK QCP, however, $Im(T_{\alpha\alpha^{\prime}}(\omega))$  
acquires an additional anomalous power-law behavior:
\begin{equation}
Im(T_{\alpha\alpha^{\prime}}(\omega))\propto \frac{1}{\omega^{-\eta_{T}^{2CK}-\eta_T}}.
\end{equation}
 where the additional anomalous exponent $\eta_T$ is 
obtained via the renormalization 
factor $Z_T$ for the $T-$matrix propagator $T_{\alpha\alpha^{\prime}}(\omega)$\cite{kircan,QMSi}: $\eta_T = \beta(j_{2}^{xy}) \frac{\partial \ln Z_T}{\partial 
j_{2}^{xy}}|_{j_{2c}^{xy}, j_{1c}^z} + \beta(j_{1}^{z}) \frac{\partial \ln Z_T}{\partial j_{1}^{z}}|_{j_{2c}^{xy},j_{1c}^{z}}.$ 
Here, the renormalization factor $Z_T$ is obtained by minimal subtraction 
of poles\cite{kircan,QMSi}: $Z_T= \frac{Z_{f}^2}{Z_{j^{\perp}}^2}$ with $Z_f$, $Z_{j^{\perp}}$ given by Eq.~\ref{Z-factors}. 
We find therefore
\begin{equation} 
\eta_T = 
\frac{(J_{2c}^{xy})^2}{2} -2 j_{1c}^z + \frac{(j_{1c}^z)^2}{4}= \frac{\epsilon}{4K}-2\epsilon + \frac{\epsilon^2}{4},
\end{equation} 
and $Im(T_{\alpha\alpha^{\prime}}(\omega))$ at QCP behaves as:  
\begin{equation}
Im(T_{\alpha\alpha^{\prime}}(\omega))\propto 
\omega^{\frac{\epsilon}{4K}+ \frac{\epsilon^2}{4}}.
\end{equation}

\subsubsection{\it Local spin susceptibility $\chi_{zz}(\omega)$.}
The local dynamical spin susceptibility $Im(\chi_{zz}(\omega))$ 
at the impurity (quantum dot) is defined as the time Fourier transform 
of the spin-spin correlator: $<S_z(0)S_z(t)>$. At zero temperature, the  
imaginary part of the local susceptibility, $Im(\chi_{zz}(\omega))$, 
shows a power-law behavior at QCP: 
\begin{equation}
Im(\chi(\omega)_{QCP})\propto \frac{1}{\omega^{-\eta_{\chi}^{2CK}-
\eta_{\chi}}}. 
\end{equation}
Here, $\eta_{\chi}^{2CK}= \epsilon$ 
 is the anomalous exponent of $Im(\chi_{zz}(\omega))$ at the tree level 
with respect to the 2CK fixed point 
via the correlator 
$<S_z(0)S_z(t)>\propto \frac{1}{t^{\frac{1}{K}}}$ evaluated at 
2CK fixed point\cite{TKNg} 
({\it ie.}, $Im(\chi_{zz}^{2CK}(\omega))\propto \omega^{\epsilon}$), 
and $\eta_{\chi}$ is the 
correction to the anomalous exponent $\eta_{\chi}^{2CK}$ 
when the system is at QCP. 
Via $\epsilon-$expansion within the field-theoretical RG 
framework\cite{kircan,QMSi}, $\eta_{\chi}$ reads:   
$\eta_{\chi} = \beta(j_{2}^{xy}) 
\frac{\partial \ln Z_{\chi}}{\partial j_{2}^{xy}}|_{j_{2c}^{xy},j_{1c}^z} +  
\beta(j_{1}^{z}) 
\frac{\partial \ln Z_{\chi}}{\partial j_{1}^{z}}|_{j_{2c}^{xy},j_{1c}^z}$ with 
$Z_{\chi}= Z_f^2$  being the renormalization factor for the impurity susceptibility\cite{kircan,QMSi} and $Z_f$ defined in Eq.~\ref{Z-factors}. 
Carrying out the above calculations, we arrive at 
\begin{equation}
\eta_{\chi} = \frac{(j_{2c}^{xy})^2}{2} + \frac{(j_{1c}^{z})^2}{4}= 
\frac{\epsilon}{4K} + \frac{\epsilon^2}{4},
\label{eta-chi}
\end{equation} 
and finally 
$Im(\chi_{zz}^{2CK}(\omega))$ at QCP shows the following power-law behaviors: 
\begin{equation}
Im(\chi_{zz}(\omega))\propto \omega^{\epsilon + \frac{\epsilon}{4K} + \frac{\epsilon^2}{4}}.
\end{equation}

\begin{figure}[t]
\begin{center}
\vspace{0.2cm}
\includegraphics[width=6.5cm]{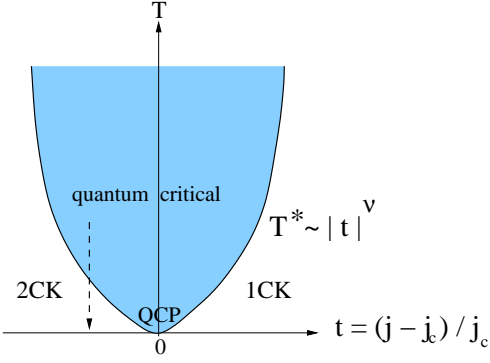}
\end{center}
\par
 \vskip -0.5cm
\caption{
(Color online)
 Schematic finite temperature phase diagram of our model near 1CK-2CK 
quantum critical point (QCP) located at $t=0$ ($j_0=j_c$). 
As $T\rightarrow 0$, the 1CK ground state is 
reached when $t>0$ or $j_0>j_c$; while as the 2CK ground state is reached 
for $t<0$ or $j_0<j_c$. Here, $j_0>(<)j_c$ refers to the 
upper right (lower left) 
region in the phase diagram shown in Fig.~\ref{phasedia} where the RG flows 
are towards the 1CK (2CK) fixed point. 
The dashed vertical arrow at $t<0$ 
refers to the finite temperature crossover 
between the quantum critical region (blue shaded area bounded by the 
crossover temperature $T^{\ast}$) and the 2CK ground state, 
which is our interest. Here, $j_0$ and $j_c$ are defined in the text.
}
\label{QCP}
\end{figure}

\subsection{\bf Hyperscaling.}
The impurity correlations at QCP is expected to obey certain hyperscaling 
properties. For example, the local dynamic spin susceptibility at criticality 
obeys $\frac{\omega}{T}$ scaling in the following form:
\begin{equation}
Im(\chi_{loc}(\omega,T)) = 
\frac{\mathcal{A}}{\omega^{-\eta_{\chi}^{2CK}-\eta_{\chi} }} \Phi(\frac{\omega}{T})
\end{equation}
with $\Phi(\frac{\omega}{T})$ being an universal crossover function for 
the QCP here and $\mathcal{A}$ being a non-universal pre-factor. 
Similar scaling form can be found in the T-matrix.  
Hyperscaling can be used to determine relations between various 
critical exponents. It has been known\cite{ingersent,kircan,QMSi,lars} that 
the correlation length exponent $\nu$ and the anomalous exponent 
$\eta_{\chi}$ are sufficient to determine all critical exponents associated 
with a local field $h$. In particular, the exponents $\gamma$ and 
$\gamma^{\prime}$ via the $T\rightarrow 0$ limit of the local susceptibility 
near criticality are defined as\cite{ingersent,lars}:
\begin{eqnarray}
\chi_{loc}(t<0;T=0) &\propto& (-t)^{-\gamma}, \;\;\; \gamma = \nu (1-\eta_{\chi}),\nonumber \\
T\chi_{loc}(t>0;T=0) &\propto& t^{\gamma^{\prime}}, \;\;\; \gamma^{\prime} = \nu \eta_{\chi}.
\label{gamma-gammap}
\end{eqnarray}
Meanwhile, the critical exponents $\beta$ and $\delta$ associated 
with the local magnetization $m_{loc}$ can be determined by:
\begin{eqnarray}
m_{loc}(t>0;T=0) &\propto& t^{\beta}, \;\;\; \beta = \frac{1}{2}\nu \eta_{\chi},\nonumber \\
m_{loc}(t=0,T=0) &\propto& |h|^{\frac{1}{\delta}}, \;\;\; \delta = 
\frac{2}{\eta_{\chi}} -1.
\end{eqnarray}
With the values for critical exponents $\nu$ (Eq.~\ref{exponent-nu}) 
and $\eta_{\chi}$ (Eq.~\ref{eta-chi}) at hand, the other critical 
exponents are therefore given by:
\begin{eqnarray}
\gamma &=& \frac{1}{2\epsilon} - \frac{1+K\epsilon}{8K} + \mathcal{O}(\epsilon^2,\epsilon^{\prime 2}), \nonumber \\
\gamma^{\prime} &=& \frac{1+K \epsilon}{8K} + \mathcal{O}(\epsilon^2,\epsilon^{\prime 2}),\nonumber \\
\beta &=& \frac{1+K \epsilon}{16K} + \mathcal{O}(\epsilon^2,\epsilon^{\prime 2}),\nonumber \\
\delta &=& \frac{8K}{\epsilon + K\epsilon^2} -1 + \mathcal{O}(\epsilon^2,\epsilon^{\prime 2}).
\end{eqnarray}

\subsection{\bf Crossover near critical point.} 
Next, we focus on calculating the crossover functions close to the 1CK-2CK 
quantum critical point. 
In general, the crossover functions of observables near criticality 
depend on the RG flows of both $j_1^z$ and $j_2^{xy}$ 
(see Fig.~\ref{QCP}); therefore they may not 
be expressed analytically in terms of universal crossover functions of a 
single variable. 
Nevertheless, great progress can be made when one makes a special 
choice of bare (initial) values of Kondo couplings such that 
$\epsilon^{\prime} j_{1,0}^{z}= (j_{2,0}^{xy})^2$. Note that this set of 
bare couplings can in general be tuned through adjusting various 
microscopic parameters, such as: spin-orbit coupling in 2DTIs, 
the lead-dot hoping. 
For this particular choice of 
bare couplings, we found and checked 
that the RG flows of $j_1^z(\mu)$ and $j_2^{xy}(\mu)$ 
follow the well-approximated trajectory: 
$\epsilon^{\prime}j_{1}^{z} \approx (j_{2}^{xy})^2$, 
({\it i.e.}, $\beta(j_{1}^z) \approx 0$). 
Under this constraint, only one  
RG $\beta-$function ($\beta(j_2^{xy})$) effectively remains: 
\begin{equation}
\beta(j_{2}^{xy}) = \epsilon j_{2}^{xy} - 2K (j_{2}^{xy})^3. 
\label{RGj2}
\end{equation}
One can therefore easily solve Eq.~\ref{RGj2} analytically, and its solution 
for the range between 
QCP at $j_{2c}^{xy} = \sqrt{\frac{\epsilon}{2K}} = 
\sqrt{\epsilon\epsilon^{\prime}}$ and the 2CK fixed point  
($j_{2}^{xy}< j_{2c}^{xy}$ where our RG and 
$\epsilon$-expansion approach is controlled) is found to be: 
\begin{equation}
j_{2}^{xy} (\mu) =  
\frac{j_{2c}^{xy}}{\sqrt{1+ (\frac{\mu}{T^*})^{-2\epsilon}}}. 
\label{j2_QCP}
\end{equation} 
where $T^* = (\frac{(j_{2c}^{xy})^2 - (j_{2,0}^{xy})^2}{ (j_{2,0}^{xy})^2})^{\frac{1}{2\epsilon}}$ is the crossover energy scale. It is clear that the power-law vanish 
of $T^{\ast}$ follows: $T^{\ast}\propto |t|^{\frac{1}{2\epsilon}}\equiv  
|t|^{\nu}$ with the correlation length exponent $\nu$ being  
$\nu= \frac{1}{2\epsilon}$, which agrees 
with our earlier result in Eq.~\ref{exponent-nu}. 
The crossover function in Eq. \ref{j2_QCP} can be used to compute various 
crossovers in thermodynamic functions near 1CK-2CK QCP as discussed below. 

\begin{figure}[t]
\begin{center}
\includegraphics[width=8cm]{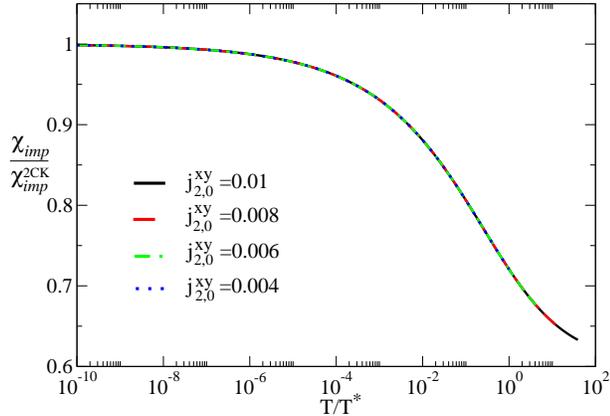}
\end{center}
\par
 \vskip -0.7cm
\caption{
(Color online)
$\chi_{imp}/\chi_{imp}^{2CK}$ versus $T/T^{\ast}$ (see Eq.\ref{chiimp}) 
at fixed $K=0.8$ for various bare Kondo couplings (in units of $\mu_0=1$) 
where $j_{1}^z = 2 K (j_2^{xy})^2$ is satisfied and 
$j_{2c}^{xy}= \sqrt{\frac{\epsilon}{2K}}\approx 0.39$. 
}
\label{chiT}
\end{figure}

\subsubsection{\it The impurity susceptibility $T\chi_{imp}(T)$.} 
 The impurity susceptibility is defined as\cite{kircan,lars}: 
$\chi_{imp} (T)= \chi_{imp,imp}+2\chi_{u,imp}+(\chi_{u,u}-\chi^{bulk}_{u,u})$
 where $\chi_{u,u}$ is the bulk response to the local field applied to the 
bulk only, $\chi_{imp,imp}$ is the impurity response to the local filed 
applied to the impurity only, $\chi_{u,imp}$ is the crossed response 
of the bulk to an impurity field, $\chi^{bulk}_{u,u}$ is the susceptibility 
of the bulk in the absence of the impurity. We can calculate 
$\chi_{imp}(T)$ via perturbative approaches 
in Refs.\cite{kircan,lars}.  
We find (up to the first order in $j_2^{xy}$) $\chi_{imp}(T)$ has the following 
crossover form (see Eq. \ref{j2_QCP} and Fig.~\ref{chiT}): 
\begin{eqnarray}
\frac{\chi_{imp}(T)}{\chi_{imp}^{2CK}(T)} &\approx& 1- j_{2}^{xy}(\mu\rightarrow T)\nonumber \\
&\approx& 1- \frac{j_{2c}^{xy}}{\sqrt{1+ (\frac{T}{T^*})^{-2\epsilon}}} 
\label{chiimp}
\end{eqnarray} 
 where $\chi_{imp}^{2CK}$ is the impurity susceptibility 
at the 2CK fixed point, given by\cite{bosonization}  
$\chi_{imp}^{2CK}(T)\propto \frac{\partial C_{imp}^{2CK}}{\partial T}
\propto \frac{1}{T^{1-\eta_{\chi_{imp}}^{2CK}}}$ with impurity specific heat 
at 2CK fixed point given by:   
$C_{imp}^{2CK} \propto T^{\frac{2}{K} -2}$ (for $\frac{2}{3}<K<1$) 
and $C_{imp}^{2CK} \propto T$ (for $K<\frac{2}{3}$)\cite{TKNg,gogolin2}. 
We have therefore $\eta_{\chi_{imp}}^{2CK}= \frac{2}{K}-2$ (for $\frac{2}{3}<K<1$) 
and $\eta_{\chi_{imp}}^{2CK}= 1$ (for $\frac{2}{3}<K<1$).\\ 
\subsubsection{\it Impurity entropy $S_{imp}(T)$.} 
At 2CK fixed point, 
the impurity residual 
entropy has been calculated in Ref.\cite{TKNg}: $S_{imp}^{2CK} = \ln\sqrt{2K}$. 
Following Ref.\cite{kircan}, the 
correction to $S_{imp}^{2CK}$ near QCP is obtained 
within perturbative RG approach the by calculating the 
thermodynamic potential and taking 
the temperature derivative. The crossover function for the 
impurity entropy near QCP is found to 
be\cite{kircan}: 
\begin{eqnarray}
\frac{S_{imp}(T)}{S_{imp}^{2CK}} &\approx& 1+ 
\frac{\pi^2 \epsilon}{4}\frac{\ln 2}{\ln \sqrt{2K}} 
[j_{2}^{xy}(\mu\rightarrow T)]^2\nonumber \\
&\approx &  1+ 
\frac{\pi^2 \epsilon}{4}\frac{\ln 2}{\ln \sqrt{2K}}
[\frac{j_{2c}^{xy}}{\sqrt{1+ (\frac{T}{T^*})^{-2\epsilon}}}]^2.
\end{eqnarray} 
\begin{figure}[t]
\begin{center}
\includegraphics[width=8cm]{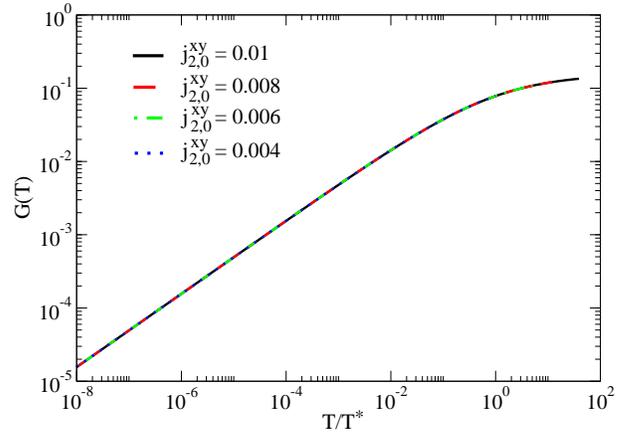}
\end{center}
\par
 \vskip -0.7cm
\caption{
(Color online)
 Crossover of the linear conductance $G(T)$ 
versus $T/T^{\ast}$ (see Eq.\ref{GT}) at fixed $K=0.8$ 
for various bare Kondo couplings (same as in Fig.\ref{chiT}).
}
\label{G(T)}
\end{figure}
\subsubsection{\it Equilibrium conductance $G(T)$.}
The equilibrium conductance $G(T)$ has the following 
crossover form between 2CK fixed point and the 
QCP (see Fig.~\ref{G(T)}): 
\begin{equation}
G(T)\propto [j_{2}^{xy}(\mu\rightarrow T)]^2
    \approx \frac{[j_{2c}^{xy}]^2 T^{2\epsilon}}
{T^{2\epsilon}+ (T^{\ast})^{2\epsilon}}.
\label{GT}
\end{equation}
Note that in equilibrium the linear conductance at 2CK fixed point 
$G_{2CK}(T)$ is determined by the bare scaling dimension of the 
leading irrelevant operator $j_2^{xy}$, $[j_2^{xy}]= \frac{1}{K}$. 
This gives $G_{2CK}(T)\propto T^{2(\frac{1}{K}-1)} = T^{2\epsilon}$. 
For $T\ll T^{\ast}$ where 
the system reaches the 2CK fixed point, the temperature dependence 
of $G(T)$ in Eq.~\ref{GT} reduces to that at 2CK, 
$G(T\ll T^{\ast})\propto G_{2CK}(T)$, as expected.

\section{Discussions and Conclusions.} 
Before we conclude, we would like to emphasize again the clear physical picture 
we provided in the Introduction to make our main results more transparent. 
First, it is well-known that the stable one-channel and two-channel 
Kondo fixed points
are expected in the case of Kondo quantum dot coupled
to two conventional spinful Luttinger liquid leads\cite{gogolin,kim}.
The ground state of this system changes from
1CK to 2CK when Luttinger parameter $K$ reduces from
the non-interacting limit ($K=1$) to the strongly interacting
limit ($K<\frac{1}{2}$). The electron-electron interactions in the
Luttinger liquids act equivalently 
as if an additional dissipative Ohmic boson bath is 
coupled to the quantum dot\cite{hur,florens}, 
leading to suppression in electron transport
from one lead to the other through the dot. A quantum phase transition
between 1CK and 2CK fixed points was argued to exist at $K=\frac{1}{2}$  
as a direct consequence of the competition between 
the cross-channel Kondo coupling $J_{LR}$ and the suppression of tunneling
due to electron-electron interaction\cite{gogolin,kim}. However, 
up to now there is no analytic and controlled approach
to access this transition. Note that the 1-loop RG approach does not work
here to reach to the 1CK-2CK quantum critical point since 
near 2CK fixed point the Kondo couplings $J_{LL/RR}$, involving in the 
renormalization of the cross-channel Kondo coupling $J_{LR}$, 
both go to infinity under RG.

 When a quantum dot couples to 
helical Luttinger liquids (a special type of Luttinger
liquid), we expect the 1CK and 2CK ground states
are also the two possible stable phases for the same reason mentioned above. 
However, due to the helical nature of the Luttinger liquid leads, 
the underling two-channel Kondo model becomes anisotropic 
($J_i^{xy}\neq J_i^z$) as the $SU(2)$ symmetry of the model is broken; 
while as the Kondo model is isotropic ($J_i^{xy}= J_i^z$)
for a quantum dot coupled to conventional Luttinger liquid leads. 
This crucial difference enables us to access the QPT between 
1CK and 2CK fixed point of our system via the controlled RG approach. 
  
In the limit of a weakly interacting helical liquid $K\rightarrow 1^{-}$, 
we find the similar competition between these two possible ground states.
The 1CK phase is reached when $J_{2}^{xy}$ is large enough; while as the 2CK
phase is reached when the electron-electron 
interaction becomes strong enough. Via a controlled perturbative RG 
approach at 1-loop order, we find that 
the 1CK-2CK quantum phase transition occurs near $K=1^{-}$. 
To reach the 1CK-2CK phase transition in our setup, 
we believe it is necessary to go beyond
the tree-level bare scaling dimension analysis, which 
predicts a stable 2CK phase for as
long as $K<1$\cite{TKNg}. The 1-loop RG is the leading correction to the 
above-mentioned bare scaling dimension analysis. 
Note that the main difference between the case for conventional Luttinger
liquid and that for helical liquid is that the resulting two-channel Kondo
model is isotropic in the former case; while it is anisotropic in the 
latter case. This difference affects details of the critical properties, 
such as: the critical points occurs at $K=K_c = \frac{1}{2}$ for the Kondo 
dot coupled to Luttinger liquid; while as $K_c = 1- \mathcal{O}(\epsilon)$ 
for the case of helical Luttinger liquid. 
At a general level, however, we should expect a 
1CK-2CK quantum phase transition to exist in both cases.

In summary, we have re-examined Ref.~\cite{TKNg} on 
the two-channel Kondo physics 
in the Kondo quantum dot coupled to 
two helical edge states of 2-dimensional topological insulators.  
Via the 1-loop renormalization group approach which goes beyond 
the scaling dimension analysis in Ref. \cite{TKNg}, we found the 
quantum phase transition between the one-channel (1CK) 
and two-channel (2CK) Kondo 
ground states for weakly interacting leads ($K\rightarrow 1^-$).   
We made definite predictions on the critical properties when the 
system is close to the transition. Our results are robust 
for $\frac{1}{2}<K<1$, and they refine the statement in Ref.~\cite{TKNg} that 
the two-channel Kondo ground state is stable for as long as $K<1$. Our results 
also provide the first theoretical realization of the 
quantum phase transition between 1CK and 2CK physics in Kondo 
impurity models. Further investigations via field-theoretical 
and Numerical Renormalization Group (NRG)\cite{hewson} approaches 
are needed in order to clarify 
the critical properties, including the critical exponents and 
finite-temperature dynamics in crossover functions associated 
with the transition\cite{lars-florens}. 
Our results could in principle motivate the search for these critical 
properties near 1CK-2CK quantum phase transition in future experiments 
on Kondo quantum dot coupled to 2D topological insulators. 



\acknowledgements
We thank  M. Vojta, T.K. Ng, K.T. Law, Y.W. Li, H.H. Lin 
for helpful discussions; and thank T.H. Lee for technical support.
This work is supported by the NSC grant
No.98-2918-I-009-06, No.98-2112-M-009-010-MY3, the NCTU-CTS, 
the MOE-ATU program, the NCTS of Taiwan, R.O.C..


\appendix 

\section{The RG scaling equation in the weak-coupling regime}

In this Appendix, we provide some details on deriving the 
RG scaling equations of Eq.~\ref{RGweak} 
for $K\rightarrow 1^-$ from 
the bosonized Hamiltonian Eq.~\ref{H_weak_bosonized}. 
Following Refs.~\cite{bosonizedRG,bosonization}, 
we decompose the boson fields $\Phi_\nu\equiv \theta_\nu, \phi_\nu$ 
with $\nu=s,a$ into the ``fast'' ($\Phi_\nu^>$) 
and ``slow'' ($\Phi_\nu^<$) components:
\begin{eqnarray}
\Phi_\nu(\tau) &=& \Phi_\nu^<(\tau) + \Phi_\nu^>(\tau),\nonumber \\
 \Phi_\nu^<(\tau) &=& \frac{1}{\beta} \sum_{|\omega_n|<\mu^\prime} 
\Phi(\omega_n),\nonumber \\
\Phi_\nu^>(\tau) &=& \frac{1}{\beta} 
\sum_{\mu^\prime <|\omega_n|<\mu}\Phi(\omega_n)\nonumber \\
\label{Phi-slow-fast}
\end{eqnarray}
with $\mu^\prime = \mu + d\mu$. 
The partition function can be decomposed in the following form:
\begin{eqnarray}
Z_\mu &=& \int D\Phi^< D\Phi^> e^{-S_0[\Phi^<]- S_0[\Phi^>] - 
S_{int}[\Phi^>+\Phi^<]},\nonumber \\
&=& Z_0 \int D\Phi^< e^{-S_0[\Phi^<]} 
<e^{-S_{int}[\Phi^<+\Phi^>]}>_{f}\nonumber \\
\label{Z}
\end{eqnarray}
where
\begin{eqnarray}
Z_0 &\equiv& \int D\Phi^> e^{-S_0[\Phi^>]},\nonumber \\
<\mathcal{A}>_{f} &\equiv& \int D\Phi^> e^{-S_0[\Phi^>]} \mathcal{A}[\Phi^>].
\end{eqnarray}
The partition function $Z_\mu$ can be re-expressed by exponentiating 
$<...>_{f}$ in the integrand in terms of the effective action 
$S_{eff}[\Phi^<]\equiv \int d\tau \mathcal{L}_K(\Phi^<)$ 
with $\mathcal{L}_K$ being the Lagrangian of the Kondo 
model (see Eq.~\ref{H_weak_bosonized}), 
involving only the slow component of the fields 
with the following form via the cummulant expansion:
\begin{eqnarray}
S_{eff}[\Phi^<] &=& S_0[\Phi^<] - \ln<e^{-S_{int}[\Phi^<+\Phi^>]}>,\nonumber \\
&=& S_0[\Phi^<]\nonumber \\
&+& <S_{int}[\Phi^<+\Phi^>]>_{f}\nonumber \\
 &-& \frac{1}{2} (<S_{int}^2[\Phi^<+\Phi^>]\nonumber \\
&-&<S_{int}[\Phi^<+\Phi^>]>^2_{f}>)+\cdots.\nonumber \\
\end{eqnarray} 
The RG procedure is carried out by integrating out the fast modes of bosons 
and expressing the effective low-energy theory in the original form 
with the renormalized couplings. The following two-point correlation functions 
of boson fields prove to be useful in the RG analysis\cite{bosonizedRG}:
\begin{eqnarray}
G(x,\tau) &=& <\Phi(x,\tau) \Phi(0,0)>_{f} \nonumber\\
&=& \int \frac{dk}{2\pi} \int \frac{d\omega}{2\pi} 
e^{-{\it i} k x} e^{{\it i} \omega \tau} 
\frac{\pi}{\frac{\omega^2}{v_F^\prime}+ v_F^\prime k^2},\nonumber \\
G(\tau) &\equiv& G(0,\tau) = 
\left\{
            \begin{array}{ll}
                  \frac{1}{2\pi}K_0(\mu^\prime\tau) & 
\mbox{for $\mu^\prime\tau >> 1 $} \\
                   \frac{1}{2\pi}\ln{\frac{\mu}{\mu^\prime}} & 
\mbox{for $\mu^\prime\tau << 1,$}
           \end{array}
            \right. \nonumber \\
\label{Gtau}
\end{eqnarray} 
where $K_0$ is the Bessel function of the second kind. It is clear from 
Eq.~\ref{Gtau} that $G(\tau)$ can be 
considered a short-ranged function of $\tau$.

First, we focus on the first order cummulant  
$<S_{int}[\Phi^>+\Phi^<]>$, which leads to the bare scaling dimensions of 
various Kondo couplings in Ref.~\cite{TKNg}. 
The renormalization of the forward longitudinal term $J_1^z$ term, 
$\delta J_1^z$, gives:
\begin{eqnarray}
&J_1^z& \int d\tau <\partial_x\theta_s(0,\tau)>_{f}\nonumber \\ 
&=& J_1^z \int d\tau [\partial_x\theta_s^{<}(0,\tau) + 
<\partial_x\theta_s^{>}(0,\tau)>_{f}].
\end{eqnarray}
Since $\theta_s$ is an odd function in spin space, its average vanishes, 
$<\partial_x\theta_s^{>}(0,\tau)>_f =0$, $\delta J_1^z=0$. 
This gives the first-order 
RG scaling equation:
\begin{equation} 
\frac{d j_1^z}{d\ln \mu}=0 
\end{equation}
with the renormalized dimensionless coupling $j_1^z$ defined as: 
$j_1^z = \rho_0 J_1^z$ where $\rho_0=\frac{1}{\pi v_F^\prime}$ is the density 
of states. The rescaling of backward longitudinal term $J_2^z$ term leads to:
\begin{eqnarray}
&\int& d\tau J_2^z <\sin(\sqrt{\frac{2\pi}{K}}\theta_a(0,\tau))>_{f} 
<\sin(\sqrt{2\pi K}\phi_a(0,\tau))>_{f} \nonumber \\
&=& (\frac{\mu^\prime}{\mu})^{\frac{K}{2}+\frac{1}{2K}} \times \nonumber \\ 
&\int& d\tau J_2^z 
\sin(\sqrt{\frac{2\pi}{K}}\theta_s^<(0,\tau)) 
\sin(\sqrt{2\pi K}\phi_s^<(0,\tau))
\label{j2z-bosonized}
\end{eqnarray} 
where Eq.~\ref{Gtau} and 
$<e^{\mathcal{A}}> = e^{\frac{1}{2} <\mathcal{A}^2>}$ are used\cite{bosonization}. 
Upon rescaling $\tau$, $\tau\rightarrow \tau \frac{\mu}{\mu^\prime}$, 
we may define the new dimensionless renormalized coupling $j_2^z(\mu)$ 
in terms of the bare coupling $J_2^z(\mu=\mu_0=1)$ as:
\begin{equation}
j_2^z(\mu) = \rho_0 \mu^{\frac{K}{2}+\frac{1}{2K}-1} J_2^z, 
\end{equation}
we arrives at the RG scaling equation at the level of bare scaling dimension:
\begin{equation}
\frac{d j_2^z}{d\ln \mu} = (\frac{K+1/K}{2} -1) j_2^z.
\end{equation}
The first-order RG scaling equations for the remaining couplings 
are obtained similarly:
\begin{eqnarray}
\frac{d j_1^{xy}}{d\ln \mu} &=& (K-1) j_1^{xy},\nonumber \\
\frac{d j_2^{xy}}{d\ln \mu } &=& (\frac{K+1/K}{2}-1) j_2^{xy}
\end{eqnarray}
with the renormalized dimensionless couplings defined in the text. 
Next, we consider the second order cummulant terms generated from 
$-\frac{1}{2} (<S_{int}^2[\Phi^<+\Phi^>] -<S_{int}[\Phi^<+\Phi^>]>^2>_f)$. 
In general, the second-order contributions to 
the renormalization of various couplings 
have the following form:
\begin{eqnarray}
\frac{d j_1^{xy}}{d\ln \mu} &=& -a_1 j_1^{xy} j_1^z - a_2 j_2^{xy} j_2^z,\nonumber \\
\frac{d j_1^z}{d\ln \mu} &=& -b_1 (j_1^{xy})^2 - b_2 (j_2^{xy})^2,\nonumber \\
\frac{d j_2^{xy}}{d\ln \mu} &=& -c_1 j_1^{xy} j_2^z - c_2 j_2^{xy} j_1^z,\nonumber \\
\frac{d j_2^z}{d\ln \mu} &=& -2 d_1 j_1^{xy} j_2^{xy},\nonumber \\
\label{RG-j1j2-bosonized}
\end{eqnarray}
with $a_i$, $b_i$, $c_i$, and $d_i$ being the pre-factors to be determined. 

We first focus on the terms in $J_2^{xy} J_2^z$ which will 
contribute to the renormalization of $j_1^{xy}$: 
\begin{eqnarray}
 & & \frac{2 J_2^{xy}J_2^z}{(\pi a)^2} S_z S^{+} \int d\tau \int d\tau^\prime 
\nonumber \\ 
&\times& 2 [ <e^{-{\it i}\sqrt{2\pi K}\phi_s(0,\tau)} 
\cos(\sqrt{\frac{2\pi}{K}}\theta_a(0,\tau)) \nonumber \\ 
&\times& \sin(\sqrt{\frac{2\pi}{K}}\theta_a(0,\tau^\prime)) 
\sin(\sqrt{2\pi K}\phi_a(0,\tau^\prime))>_f \nonumber \\
&-&  <e^{-{\it i}\sqrt{2\pi K}\phi_s(0,\tau)} 
\cos(\sqrt{\frac{2\pi}{K}}\theta_a(0,\tau))>_f \nonumber \\ 
&\times& <\sin(\sqrt{\frac{2\pi}{K}}\theta_a(0,\tau^\prime)) 
\sin(\sqrt{2\pi K}\phi_a(0,\tau^\prime))>_f ]. \nonumber \\
\end{eqnarray}
After averaging over the fast modes and rescaling $\tau ,\tau^\prime$, 
we arrive at:
\begin{eqnarray}
& & \frac{J_2^{xy}J_2^z }{(\pi a)^2} S^+\int d\tau \int d\tau^\prime 
(\frac{\mu^\prime}{\mu})^{\frac{1}{K}+K-2}\nonumber \\ 
&\times & ((\frac{\mu^\prime}{\mu})^{\frac{-1}{K}} -1) 
e^{-{\it i}\sqrt{2\pi K}\phi_s^<(0,\tau)} 
\cos(\sqrt{2\pi K}\phi_a^<(0,\tau^\prime)),\nonumber \\ 
\label{j2pj2z-bosonized}
\end{eqnarray}
In deriving the above equation, we have decomposed the terms 
$\sin(\sqrt{2\pi K}(\phi_a(\tau^\prime))$ 
and $\cos(\sqrt{\frac{2\pi}{K}}(\theta_a(\tau)) 
\sin(\sqrt{\frac{2\pi}{K}}(\theta_a(\tau^\prime))$ into the fast and the 
slow modes, and kept only the leading (more relevant) terms. 
In the limit of $\tau,\tau^\prime\ll \frac{1}{\mu}\approx a$, 
we may get rid off one of the double time-integrals in the above 
equation by introducing a short-time cutoff 
$\tau_0\approx \frac{a}{v_F^\prime}$. 
In the limit of $K\rightarrow 1^-$, Eq.~\ref{j2pj2z-bosonized} becomes:
\begin{equation}
-\frac{j_2^{xy}j_2^z }{\pi a} S^+\int d\tau  
\frac{d\mu}{\mu}
e^{-{\it i}\sqrt{2\pi K}\phi_s^<(0,\tau)} 
\cos(\sqrt{2\pi K}\phi_a^<(0,\tau)).
\label{j2pj2z-final}
\end{equation}
Therefore, the pre-factor $a_2$ in Eq.~\ref{RG-j1j2-bosonized} 
is found to be $a_2=1$. Similarly, we find the pre-factors 
$c_1=b_1=b_2=d_1=1$ in Eq.~\ref{RG-j1j2-bosonized}.

Next, we consider a different type of renormalization involving $J_1^z$ terms. 
We may focus on a typical term $J_1^{xy} J_1^z$, which renormalizes $J_1^{xy}$:
\begin{eqnarray}
&-&a\sqrt{\frac{2\pi}{K}}\frac{J_1^{xy}J_1^{z}}{(\pi a)^2} 
\int d\tau \int d\tau^\prime S^{-} S_z \nonumber \\
&\times&[<\partial_x\theta_s(0,\tau) \nonumber \\
&\times& e^{-{\it i}\sqrt{2\pi K}\phi_s(0,\tau^\prime)} 
\cos(\sqrt{2\pi K}\phi_a(0,\tau^\prime))>_f \nonumber \\ 
&-&
<\partial_x\theta_s(0,\tau)>_f \nonumber \\
&\times& < e^{-{\it i}\sqrt{2\pi K}\phi_s(0,\tau^\prime)} 
\cos(\sqrt{2\pi K}\phi_a(0,\tau^\prime))>_f ]\nonumber\\
&=& -a\sqrt{\frac{2\pi}{K}}\frac{J_1^{xy}J_1^{z}}{(\pi a)^2} 
\int d\tau \int d\tau^\prime S^{-} S_z \nonumber \\
&\times&[<\partial_x\theta_s^>(0,\tau)\nonumber \\
&\times& e^{-{\it i}\sqrt{2\pi K}\phi_s(0,\tau^\prime)} 
\cos(\sqrt{2\pi K}\phi_a(0,\tau^\prime))>_f]. \nonumber \\ 
\label{j1pj1z-bosonizedRG}
\end{eqnarray}
We may use the following identities\cite{bosonizedRG} to 
simplify Eq.~\ref{j1pj1z-bosonizedRG}:
\begin{eqnarray}
& &<\partial_x\sqrt{\frac{2\pi}{K}} \theta_s^>(0,\tau) 
e^{-{\it i}(\sqrt{2\pi K}\phi_s(0,\tau^\prime))} 
e^{{\it i} \sqrt{2\pi K}\phi_a(0,\tau^\prime)}>_f\nonumber \\
&=& \lim_{\eta\rightarrow 0} \frac{1}{{\it i}\eta}\partial_x 
<e^{{\it i}\eta\sqrt{\frac{2\pi}{K}}\theta_a^>(x,\tau) }
e^{-{\it i}\sqrt{2\pi K}\phi_s^>(0,\tau^\prime)}\nonumber \\ 
& & e^{{\it i}\sqrt{2\pi K}\phi_a^>(0,\tau^\prime)}>_f|_{x=0}\nonumber \\ 
&\times& e^{-{\it i}(\sqrt{2\pi K}\phi_s^<(0,\tau^\prime))}
e^{{\it i}\sqrt{2\pi K}\phi_a^<(0,\tau^\prime)},\nonumber \\
\label{identity1}
\end{eqnarray}
and 
\begin{eqnarray}
& & e^{\mathcal{A}+\mathcal{B}} = e^{\mathcal{A}} e^{\mathcal{B}} 
e^{\frac{1}{2}[\mathcal{A},\mathcal{B}]}.\nonumber \\
\label{identity2}
\end{eqnarray}
 With the above relations, Eq.~\ref{identity1} becomes:
\begin{eqnarray}
& &\lim_{\eta\rightarrow 0} \frac{1}{{\it i}\eta}\partial_x 
<e^{{\it i}\sqrt{2\pi}(\eta \sqrt{\frac{1}{K}}\theta_a^>(x,\tau) 
-  \sqrt{K} \phi_s^>(0,\tau^\prime))}>_f|_{x=0} \nonumber \\
& \times & <e^{{\it i}\sqrt{2\pi K}\phi_a^>(0,\tau^\prime)}>_f\nonumber \\ 
&\times& e^{-{\it i}\sqrt{2\pi K}\phi_s^<(0,\tau^\prime)}
e^{{\it i}\sqrt{2\pi K}\phi_a^<(0,\tau^\prime)},\nonumber \\
&=&\frac{1}{{\it i}} 
(\frac{\pi}{K} \partial_x G_{\theta^>_s(x,\tau)} + 
2\pi \partial_x <\theta^>_s(x,\tau) 
\phi_s^>(0,\tau^\prime)>_f )|_{x=0}\nonumber \\ 
&\times& <e^{{\it i}\sqrt{2\pi K}\phi_s^>(0,\tau^\prime)}>_f
<e^{{\it i}\sqrt{2\pi K}\phi_a^>(0,\tau^\prime)}>_f\nonumber \\
\label{identity2}
\end{eqnarray}
The leading logarithmic correction comes from the term  
$\partial_x <\theta_s^>(x,\tau) \phi_s^>(0,\tau^\prime)>_f$, 
which can be evaluated via the following relations\cite{bosonizedRG}:
\begin{eqnarray}
&-&\frac{{\it i}}{v_F^\prime}\frac{\partial \phi_s}{\partial \tau} 
= \frac{\partial \theta_s}{\partial x},\nonumber \\
&-&\frac{{\it i}}{v_F^\prime}\frac{\partial \theta_s}{\partial \tau} 
= \frac{\partial \phi_s}{\partial x}.\nonumber \\
\end{eqnarray}
We have therefore
\begin{eqnarray}
\partial_x <\theta_s^>(x,\tau) \phi_s^>(0,\tau^\prime)>_f &=& 
\frac{-{\it i}}{v_F^\prime} \frac{\partial}{\partial \tau}
<\theta_s^>(x,\tau) \theta_s^>(0,\tau^\prime)>_f.\nonumber \\
\end{eqnarray}
After collecting all the terms and performing re-scaling, 
Eq.~\ref{j1pj1z-bosonizedRG} becomes:
\begin{eqnarray}
& &-\frac{1}{\pi v_F^\prime} \frac{J_1^{xy}J_1^z}{\pi a} 
\int d\tau S^- (\frac{\mu^\prime}{\mu})^{K-1} \frac{d\mu}{\mu}\nonumber \\
&\times & e^{-{\it i}(\sqrt{2\pi K}\phi_s^<(0,\tau^\prime))} 
\cos(\sqrt{2\pi K}\phi_a^<(0,\tau^\prime)).\nonumber \\
\end{eqnarray}
Finally, the correction to $j_1^{xy}$ contributed from $J_1^{xy}J_1^z$, 
$\delta j_{1,j_1^{xy}j_1^z}^{xy}$, reads:
\begin{eqnarray}
\delta j_{1,j_1^{xy}j_1^z}^{xy} 
&=& -j_1^{xy} j_1^z \frac{d\mu}{\mu} \int d\tau 
e^{-{\it i}\sqrt{2\pi K}\phi_s^<(0,\tau)}\nonumber \\ 
&\times& \cos(\sqrt{2\pi K}\phi_a^<(0,\tau)).
\end{eqnarray}
We therefore find $a_1=1$. Similarly, we find $c_2=1$. 
Combining the first and second order corrections to the renormalization 
of various Kondo couplings, Eq.~\ref{RGweak} follows.

\section{The 1-loop RG scaling equations near 2CK fixed point via poor-man's scaling.}

In this Appendix, we derive the RG equations in Eq. 
\ref{RG2CK} from the effective 
Kondo Hamiltonian Eq. \ref{dH2CK} in the weak coupling regime 
via poor-man's scaling as shown in Ref.~\cite{florens}. Based on the 
scaling dimensions of various Kondo couplings in the strong couping 2CK 
regime, we take the logarithmic derivative of the proposed 
new dimensionless Kondo couplings 
$j_{2}^{xy}\equiv \rho_0 \tilde{c}_2^{\perp} \mu^{\epsilon}J_2^{xy}, 
j_1^z \equiv \rho_0  \tilde{c}_1^z \mu^{\epsilon^\prime} \tilde{J}_1^z$ 
(see text) with respect to the cutoff energy $\mu$.  

First, we focus on the RG equation for 
$j_2^{xy}$:
\begin{equation}
\frac{\partial j_{2}^{xy}}{\partial \ln \mu} = \epsilon j_2^{xy} - \mu^{\epsilon} 
\rho_0 \tilde{c}_2^{\perp} \frac{\partial J_{2}^{xy}}{\partial \ln \mu}.  
\end{equation}
The derivative of $J_2^{xy}$ w.r.t. $\ln \mu$ is given by:
\begin{equation}
\frac{\partial J_{2}^{xy}}{\partial \ln \mu}  =  
\frac{\partial}{\ln \mu} \int_{\mu_0}^{\mu} d\omega 
[J_2^{xy} \tilde{J}_1^z \frac{\rho_{2xy,1z}(\omega)}{-\omega}]. 
\label{strong-RG-j2xy}
\end{equation}
Here, $\rho_{2xy,1z}(\omega)$ is the effective electron 
density of states due to the additional phase correlations 
associated with the product of $J_{2}^{xy}$ and $\tilde{J}_1^z$ terms in 
Eq.~\ref{dH2CK}. Following Ref.~\cite{florens}, 
this is equivalent to replacing the free 
electron Green's function of the effective leads:
 $G_{\alpha,0}^{\sigma}(t) \equiv 
<\tilde{c}^{\dagger\sigma}_{\alpha}(t) \tilde{c}_{\alpha}^{\sigma}(0)>$ by 
a ``mixed'' one:
\begin{eqnarray}
\tilde{G}_{\tilde{L}}^{\uparrow (\downarrow)}(t) &\equiv& 
< \tilde{c}^{\dagger\uparrow (\downarrow)}_{\tilde{L}}(t) 
\tilde{c}_{\tilde{L}}^{\uparrow \downarrow}(0) \nonumber \\
&\times&
e^{\pm{\it i}\sqrt{4\pi (\frac{1}{K}-1)} \tilde{\theta}_a (t)} S^{\pm}(0) S_z(t)>\nonumber \\
&\approx& G_{\tilde{L},0}^{\uparrow (\downarrow)}(t) \nonumber \\
&\times & <e^{\pm{\it i} \sqrt{4\pi (\frac{1}{K}-1)} \tilde{\theta}_a (t)}> 
<S^{\pm}(0) S_z(t)>.\nonumber \\ 
\end{eqnarray}
($\tilde{G}_{\tilde{R}}^{\sigma}(t)$ can be defined similarly.) 

Therefore, $\rho_{2xy,1z}(\omega) \equiv 
\frac{-1}{\pi}\sum_\sigma Im(\tilde{G}_{\alpha}^{\sigma}(\omega))$ reads\cite{florens}: 
\begin{eqnarray}
\rho_{2xy,1z}(\omega) &=& \rho_0 \int_0^{\omega} dE P_{2\perp 1z}(E),\nonumber \\
P_{2\perp 1z}(E) &=& \frac{1}{2\pi} \int dt <\hat{O}_{2\perp 1z}(t)> 
e^{{\it i} E t}
\end{eqnarray}
where $\rho_0 = \frac{-1}{\pi} \sum_{\sigma} 
Im(G_{\alpha,0}^{\sigma}(\omega))$ is 
the constant density of states of the non-interacting 
leads, and $<\hat{O}_{2\perp 1z}(t)>$ has the following typical form:
\begin{equation}
<\hat{O}_{2\perp 1z}(t)> 
= <e^{\pm {\it i} \sqrt{4\pi (\frac{1}{K}-1)} \tilde{\theta}_a (t)}> <S^{\pm}(0) S_z(t)>. 
\end{equation}
 Since the exponential factors in $\hat{O}_{2\perp 1z}(t)$ are un-paired, it 
gives a trivial result: 
$<e^{\pm {\it i} \sqrt{4\pi (\frac{1}{K}-1)} \tilde{\theta}_a (t)}> = 1$, and hence   
it does not affect the 
renormalization of the couplings\cite{florens}. Nevertheless, $<S^{\pm}(0) S_z(t)>$ shows non-trivial correlations near\cite{TKNg}:
\begin{equation}
<S^{\pm}(0) S_z(t)> \approx 
\frac{1}{({\it i} \omega_c t)^{\epsilon^{\prime}}}
\end{equation}
with $\omega_c$ being a high-energy cutoff. 
We have therefore 
\begin{eqnarray}
\rho_{2xy,1z}(\omega) &=& \rho_0 \int_0^{\omega} dE P_{2\perp 1z}(E) = \rho_0 
\tilde{c}_2 \omega^{2\epsilon^\prime},\nonumber \\
P_{2\perp 1z}(E) &=& \frac{1}{2\pi} \int dt <\hat{O}_{2\perp 1z}(t)> 
e^{{\it i} E t} = \tilde{c}_2^\perp E^{\epsilon^\prime -1}\nonumber \\
\end{eqnarray}
with $\tilde{c}_2 
= A(\epsilon^\prime)= \frac{1}{\pi \omega_c^{\epsilon^\prime}} 
\sin(\pi \epsilon^\prime) \Gamma(1+\epsilon^\prime)$ with 
$\epsilon^\prime=\frac{1}{2 K}$ 
and $\Gamma$ being the Gamma function.

 The integral in Eq.~\ref{strong-RG-j2xy} gives:
\begin{eqnarray} 
& & \rho_0^2 \tilde{c}_2^{\perp} \mu^{\epsilon} J_2^{xy} J_1^z \int_{\mu_0}^{\mu} d\omega 
\frac{\rho_{2xy,1z}(\omega)}{-\omega}\nonumber \\
&=& -\rho_0^2 \tilde{c}_2^{\perp} A(\epsilon^\prime) 
\mu^{\epsilon} \mu^{\epsilon^\prime}
J_2^{xy} J_1^z  \ln \frac{\mu}{\mu_0}.
\end{eqnarray}
Similarly, we find the RG scaling equation for 
$j_1^z$ is given by:
\begin{eqnarray}
\frac{\partial j_{1}^{z}}{\partial \ln \mu} &=& \epsilon^\prime j_1^{z}
- \rho_0^2 \tilde{c}_1^{z} A(2\epsilon) \mu^{2\epsilon}   
(J_2^{xy})^2 \nonumber \\
\label{j1z2CK}
\end{eqnarray}
 where the effective density of states is used\cite{florens}: 
\begin{eqnarray}
\rho_{2\perp,2\perp}(\omega) &=& 
\rho_0 \int_0^{\omega} dE P_{2\perp2\perp}(E),\nonumber \\
P_{2\perp 2\perp}(E) &=& \int dt <\hat{A}(t)> e^{{\it i} E t}, \nonumber \\
<\hat{A}(t)> &=& <e^{-{\it i} \sqrt{4\pi (\frac{1}{K}-1))}\tilde\theta_a (t)} 
e^{{\it i} \sqrt{4\pi (\frac{1}{K}-1)}\tilde\theta_a (0)}> \nonumber \\
&\approx& 
\frac{1}{({\it i} \omega_c t)^{2\epsilon}}. \nonumber \\
\end{eqnarray} 
We may determine the pre-factors $\tilde{c}_{2}^{\perp}, \tilde{c}^z$ 
by the following identifications: 
$\tilde{c}_2^{\perp} = \sqrt{A(\epsilon) A(\epsilon^\prime)}$,
$\tilde{c}_1^z=A(\epsilon^\prime)$. With the above results, 
we finally arrive at RG scaling equations shown in Eq.\ref{RG2CK}.

\section{The 1-loop RG equations near 2CK fixed point via $\epsilon$-expansion technique.}
 
In this Appendix, we offer an alternative route to Eq.~\ref{RG2CK} via 
$\epsilon$-expansion technique\cite{Zinn-Justin,zarand,QMSi,kircan,lars}.  

We first derive  
the renormalization factors $Z_{j^{\perp / z}}$ and $Z_f$ shown in 
Eq. \ref{Z-factors}. 
We focus on the 1-loop renormalization of the 
dimensionless bar couplings $\tilde{j}_2^{xy} \equiv \rho_0 \tilde{c}_2^{\perp} 
\mu^{\epsilon} J_2^{xy}$, and $\tilde{j}_1^z \equiv \rho_0 \tilde{c}_1^{z}
\mu^{\epsilon^{\prime}} J_1^z$. 
Let us look at vertex renormalization of $j_1^z$ 
first\cite{QMSi,kircan}:
\begin{eqnarray}
j_1^{z} &\equiv& 
Z^{-1}_{j^z} \tilde{j}_1^z 
= \tilde{j}_1^z  [1 + \frac{(J_2^{xy})^2}{\tilde{j}_1^z} 
\int^{\mu} d\omega \frac{\rho_{\perp}(\omega)}{-\omega}] 
\label{2CK-j1}
\end{eqnarray}
 where the effective density of states reads\cite{florens}: 
\begin{eqnarray}
\rho_{\perp}(\omega) &=& \rho_0 \int_0^{\omega} dE P_{\perp}(E),\nonumber \\
P_{\perp}(E) &=& \int dt <\hat{A}(t)> e^{{\it i} E t}, \nonumber \\
<\hat{A}(t)> &=& <e^{-{\it i} \sqrt{4\pi}\tilde\theta_a (t)} 
e^{{\it i} \sqrt{4\pi}\tilde\theta_a (0)}> \approx 
\frac{1}{({\it i} \omega_c t)^{2\epsilon}}. \nonumber \\
\end{eqnarray}
From above, we find 
$P_{\perp}(E)= \tilde{c}_1 E^{2\epsilon-1}$ 
with $\tilde{c}_1 = A(2\epsilon)$ 
and the constant $A(\epsilon)$ being defined in Appendix B..

Therefore, we have 
\begin{equation}
\rho_{\perp}(\omega) = \rho_0 \tilde{c}_1 \omega^{2\epsilon}.
\label{rho-perp}
\end{equation} 
Plugging in these results into 
Eq.~\ref{2CK-j1} and via the proper identification: 
$\tilde{c}_2^{\perp}\equiv \sqrt{\tilde{c}_1}=\sqrt{A(2\epsilon)}$, 
at the leading order 
in $(j_{2}^{xy})^2/j_1^z$, $Z_{j^z}$ reads: 
\begin{equation}
Z_{j^z} = 1+ \frac{(j_2^{xy})^2/j_1^z}{2 \epsilon}.  
\end{equation} 
Similarly, we can show that 
\begin{equation}
Z_{j^{\perp}} = 1+ \frac{j_1^z}{\epsilon^{\prime}}
\end{equation}
where the following relations are used: 
\begin{eqnarray}
\rho_{z}(\omega) &=& \frac{\rho_0}{2\pi} 
\int_0^{\omega} dE P_z(E)   = \rho_0 \tilde{c}_2 
\omega^{2\epsilon^{\prime}}, \nonumber \\
P_z(E) &=& \int dt <\hat{B}(t)> e^{{\it i} E t}= 
\tilde{c}_2 E^{\epsilon^{\prime}-1},\nonumber \\
\hat{B}(t) &\equiv& <S^{\pm}(0) S_z(t)> \approx 
\frac{1}{({\it i} \omega_c t)^{\epsilon^{\prime}}}\nonumber \\
\label{rho-z}
\end{eqnarray}
with $\tilde{c}_2 = A(\epsilon^{\prime})$, and 
$\tilde{c}_1^z = \sqrt{\tilde{c}_1}/\tilde{c}_2$.

Next, we provide derivation for $Z_f$. Following Ref.~\cite{QMSi}, the self 
energy at 1-loop order (see Fig. 5(a) of Ref. \cite{QMSi}) 
leads to the following renormalization factor $Z_f$ for impurity fermion:
\begin{equation}
Z_f = 1+ \frac{(J_2^{xy})^2}{4}\int^\mu d\omega 
\frac{\rho_{\perp}(\omega)}{\omega} +  \frac{(J_1^{z})^2}{8}\int^\mu d\omega 
\frac{\rho_{z}(\omega)}{\omega}.  
\label{Zf}
\end{equation}
Plugging  Eq. \ref{rho-perp} and Eq.~\ref{rho-z} 
into Eq. \ref{Zf} and expressing results in 
terms of the dimensionless couplings  $j_2^{xy}$ and $j_1^z$, we arrive at:   
\begin{equation}
Z_f = 1+ \frac{(j_2^{xy})^2}{8\epsilon} +  
\frac{(j_1^{z})^2}{16\epsilon^{\prime}}.
\end{equation}

With $Z_f$, $Z_{j^{\perp}}$, $Z_{j^z}$ at hand, we now can reproduce  
the RG scaling equations Eq. ~\ref{RG2CK} via the 
$\beta$-function within the field-theoretical $\epsilon$-expansion 
approach : 
\begin{eqnarray}
\beta(j_1^z) &\equiv& \mu 
\frac{\partial j_1^z}{\partial \mu}|_{j_{2,0}^{xy},j^z_{1,0}},\nonumber \\
\beta(j_2^{xy}) &\equiv& \mu 
\frac{\partial j_2^{xy}}{\partial \mu}|_{j^{xy}_{2,0},j_{1,0}^z} \nonumber \\
\end{eqnarray} 
with  the relations between the bare Kondo couplings 
$j_{1,0}^z$, $j_{2,0}^{xy}$ and the renormalized ones $j_1^z$, $j_2^{xy}$ being 
$j_{2,0}^{xy}\equiv J_2^{xy} = \frac{\mu^{-\epsilon} Z_{j^{\perp}}}{Z_f} 
j_2^{xy}$, and $j_{1,0}^z\equiv \tilde{J}_1^z =  
\frac{\mu^{-\epsilon^{\prime}} Z_{j^{z}}}{Z_f} j_1^{z}$.

\end{document}